\documentclass[aps,prc,superscriptaddress,showpacs,nofootinbib,showkeys,twocolumn]{revtex4}

\usepackage{graphicx}
\usepackage{epsfig}
\usepackage{amsmath}

\newcommand{\midtilde}{\raisebox{-0.25\baselineskip}{\textasciitilde}}
\usepackage{color}
\usepackage[normalem]{ulem}  
\definecolor{green2}{rgb}{0,0.8,0}
\definecolor{orange}{rgb}{1,0.5,0}

\begin{document}

\title{Constraining supernova equations of state with equilibrium constants from heavy-ion collisions}

\author{Matthias Hempel}
\email{matthias.hempel@unibas.ch}
\affiliation{Department of Physics, University of Basel, 4056 Basel, Switzerland}
\author{Kris Hagel}
\author{Joseph Natowitz}
\affiliation{Cyclotron Institute, Texas A\&M University, College Station, Texas 77843, USA}
\author{Gerd R\"opke}
\affiliation{University of Rostock, FB Physik, Rostock, Germany}
\author{Stefan Typel}
\affiliation{GSI Helmholtzzentrum f\"ur Schwerionenforschung GmbH, D-64291 Darmstadt, Germany}

\date{\today}
\keywords{Nuclear matter equation of state, Cluster formation, Supernova
simulations, Low-density nuclear
matter, Relativistic mean-field model, Nuclear statistical equilibrium, Excluded
volume}
\pacs{21.7.Mn, 26.50.+x, 21.30.Fe, 25.70.Pq}

\begin{abstract}
Cluster formation is  a fundamental aspect of the equation of state
(EOS) of warm and dense nuclear matter such as can be found in supernovae
(SNe). Similar matter can  be studied in heavy-ion
  collisions (HIC).
We use the experimental data of Qin \textit{et al}.\ [Phys.\ Rev.\ Lett. {\bf 108}, 172701 (2012)]
to test calculations of cluster formation
and the role of in-medium modifications of cluster properties in SN EOSs.
For the comparison between theory and experiment we use chemical
equilibrium constants as the main observables. 
This reduces some of the
systematic uncertainties and allows deviations from ideal gas
behavior to be identified clearly.
In the analysis, we carefully 
account for the differences between matter in SNe and HICs.
We find that, at the lowest
densities, the experiment and all theoretical models are consistent
with the ideal gas behavior. At higher densities
ideal behavior is clearly ruled out and interaction effects
have to be considered.
The contributions of continuum correlations are of relevance in the virial expansion and remain a 
difficult problem to solve at higher densities.
We conclude that at the densities and temperatures discussed mean-field interactions of nucleons, inclusion of all relevant light clusters, and a suppression mechanism of clusters at high densities have to be incorporated in the SN EOS.

\end{abstract}

\maketitle
\section{Introduction}
Cluster formation is a fundamental aspect of the equation of state (EOS) of warm and low density nuclear matter, i.e., for temperatures of several MeV and densities below normal nuclear matter density $n_B^0\sim 0.15$~fm$^{-3}$. It has been shown in several works \cite{oconnor07, arcones08,sumiyoshi08,hempel11,fischer14}, that these conditions are realized in nature in core-collapse supernovae (SNe). There, nuclear clusters appear abundantly in the shock heated matter and in the envelope of the newly born proto neutron star. However, their role on the explosion dynamics and the subsequent cooling and compression of the proto neutron star
is not yet fully understood. In particular, their impact on the evolution of the neutrino spectra and luminosities has to be investigated further. This is important because these neutrino properties determine the nucleosynthesis conditions in the so-called neutrino-driven wind, and thus the question of whether or not a full r-process is possible in core-collapse SNe.

Fortunately, cluster formation and their in-medium effects can be probed
by heavy-ion collision (HIC) experiments. Experimental data can be
used to discriminate different SN EOSs; see, e.g.,
Refs.~\cite{natowitz10,qin2012}. From theory, the necessity to account
for medium effects on clusters has been known for a long time. Only now
do the experimental data allow a detailed 
analysis of this question. To our 
knowledge, Ref.~\cite{natowitz10} is the first work with a clear statement on density effects in the EOS of clusterized matter deduced from experiments. 

In comparisons between theory and experiment problems arise, because of differences of the state of matter in SNe compared to that in HICs. Temperatures and densities are similar, but matter in SNe can be more asymmetric, i.e., can have a lower (total) proton fraction. Even more important for our investigation, the fireball in a HIC has a finite size, limiting the maximum mass number which the clusters can have. Furthermore, matter in SNe, which can be considered as being infinite, has to be charge neutral, whereas there is a net charge in HICs. This leads to different Coulomb interactions in the two systems. 

Another problem arises because some SN EOSs make limiting assumptions for the nuclear composition. For example the EOS of Shen \textit{et al}.\ \cite{shen98} (STOS) and the EOS of Lattimer and Swesty \cite{lattimer91} (LS), which are frequently used for astrophysical applications, only consider neutrons, protons, $\alpha$ particles and a representative heavy nucleus as degrees of freedom. Regarding the representative heavy nucleus, it was shown already in Ref.~\cite{burrows84} that the so-called single nucleus approximation leads only to minor deviations for thermodynamic quantities. However, for light nuclei it is not possible to introduce only a representative nucleus, as we will also show in the present study. 

The main idea of Ref.\ \cite{qin2012} was to extract a quantity from experimental data that is robust with respect 
to  effects such as the asymmetry of the source or a limited nuclear composition, which can be difficult to control either experimentally or in the theoretical description,
 as described above. Instead of using the particle yields directly, the chemical equilibrium constant (EC) was considered. This quantity is given as a particular ratio of particle number densities $n_i$ \cite{qin2012}:
\begin{equation}
 K_c[i]=\frac{n_i}{n_n^{N_i}n_p^{Z_i}} \; , \label{eq_kc}
\end{equation}
where $N_i$ and $Z_i$ are the neutron and charge number,
  respectively of nucleus $i$. 
The quantities $n_n$ and $n_p$ are the number densities of
  neutrons and protons not bound in nuclei.

The usage of the EC reduces the influence of the problematic aspects in the model comparisons, but is still able to discriminate between different EOSs. The ECs $K_c$ are sensitive to medium modifications, in particular Pauli blocking\footnote{Light clusters are composite particles of nucleons. Thus, at large densities the light clusters do not behave as free
quasiparticles, but are influenced by the filled Fermi sea of nucleons. This effect is called Pauli blocking and leads to a
shift in the binding energies.} and the dissolution of bound states at
high densities. Compared with a nuclear statistical equilibrium
(NSE) description of ideal gases (also called the ideal mass action law), the medium effects lead to a reduction of the abundances of bound states and the mass fractions of single nucleon states are increased. Both effects reduce the value of $K_c$ so that large deviations from NSE are expected, as also verified by experiments. On the other hand, $K_c$ is less sensitive to the asymmetry of nuclear matter, and also less influenced by the abundance of other (heavy) elements, than are the abundances or mass fractions of light clusters. Therefore it is a good signature for the medium modifications of light element properties in dense matter.

A quantum statistical (QS) approach to calculate the chemical
constants has been used in Ref.~\cite{qin2012}. It was shown that the
simple description of NSE based on ideal (classical) gases is clearly
ruled out at some of the densities under consideration. Interactions
of the nucleons and clusters and their medium modifications have to be
taken into account, for example as excluded volume effects or as
quasiparticle energy shifts including Pauli blocking. On the other hand, in Ref.~\cite{qin2012}, unexpected deviations were found among the different theoretical models at low densities where the ideal NSE should be a good approximation. In the present study, we improve the calculations presented in Ref.~\cite{qin2012} and achieve a better agreement of the different theoretical descriptions at low densities. 

The general aim of the present study is to refine the comparison of the predictions of various SN EOSs with the experimental results, to further constrain the medium effects at high densities. Different theoretical approaches are considered, from excluded volume to quantum statistical calculations. We will also use the ECs as the main observables, but discuss aspects which were not addressed in Ref.~\cite{qin2012}. We investigate uncertainties and model assumptions in the experimental \textit{and} theoretical data and identify to which degree they are of relevance. In particular, we discuss the sensitivity of the chemical ECs to the asymmetry of the emitting source, to limiting assumptions for the nuclear composition, and to Coulomb corrections. 

The outline of this article is as follows: in Sec.~\ref{sec_experiment} we describe the experiment and the methods used to extract the relevant data. In Sec.~\ref{sec_definitions}, we introduce some definitions used later and discuss basic aspects of ECs. In Sec.~\ref{sec_depend}, we analyze effects of the asymmetry of the system, of Coulomb interactions, and of particle degrees of freedom on the ECs, using modifications of the EOS model of Hempel and Schaffner-Bielich (HS) \cite{hempel10}. In Sec.~\ref{sec_other}, we introduce the other EOS models considered in our work and modify them such that they fit to the conditions in HICs. Section~\ref{sec_benchmarking} contains the main results of our work, where the final comparison among the different theoretical models and the experimental data is given. In Sec.~\ref{sec_summary}, we summarize and present our conclusions. 

\section{Experimental data}
\label{sec_experiment}
The NIMROD 4$\pi$ multidetector at Texas A\&M University was used to measure cluster production in collisions of 47A MeV $^{40}$Ar with $^{112,124}$Sn and $^{64}$Zn with $^{112,124}$Sn. The yields of \textit{p} ($^1$H), \textit{d} ($^2$H), \textit{t} ($^3$H), \textit{h} ($^3$He), and $\alpha$ ($\alpha$ particles, $^4$He) were determined. The combined neutron and charged particle multiplicities were employed to select the most violent events for subsequent analysis. Double isotope yield ratios \cite{albergo,kolomiets97} were used to characterize the temperature at a particular emission time, and densities were determined with the thermal coalescence model of Mekjian \cite{mekjian77,mekjian78,hagel2000}. For comparison with other methods of density determinations, see Ref.~\cite{roepke2013}. All experimental data that we are using are taken directly from Ref.~\cite{qin2012}. For all details of the experimental setup and analysis, we refer to Ref.~\cite{qin2012}.

\begin{figure}
\begin{center}
\includegraphics[width=0.825\columnwidth]{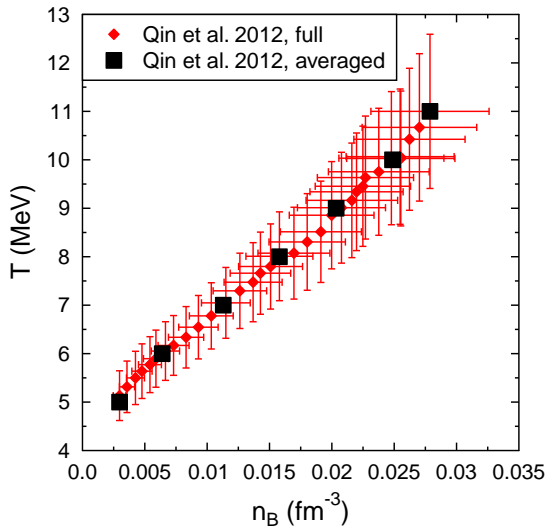}
\caption{\label{fig_tnb}(Color online) Temperatures and densities extracted from the experiment. The red diamonds show the full data set, the black squares represent a reduced average. We only show the statistical errors for the full set, because the uncertainties of the averaged one are similar.}
\end{center}
\end{figure}

The statistical uncertainty for the extracted densities are 17\%, and
for the extracted temperature are 10\% for densities
below 0.01~fm$^{-3}$,
increasing to $\sim$15\% at 0.03~fm$^{-3}$. The errors of the measured
particle yields, used to derive the values of the equilibrium
constants, are much smaller. 
Figure~\ref{fig_tnb} shows the experimental
data. There are two data sets: in addition to the standard binning,
which gives 34 different combinations of temperature and density,
there is a second data set where  the full information has been
averaged to seven data points at temperatures of 5, 6, 7, 8, 9, 10, and 11
MeV.
In the calculations we always use only the mean values of the measured temperatures and densities, and include their uncertainties in the figures by grey bands or error bars.

In the following we discuss aspects of the experiments, which are of particular importance for the present investigation. The first is the question of which nuclear clusters can be formed. The experimental data which we are using are extracted for a specific source which is identified as the early emitted gas---it leaves the system in about 150~fm/c and is a relatively small fraction ($\sim 20 \%$) of the total system, containing roughly 30 to 40 nucleons. Its momentum sphere is initially detached from that of the surrounding matter. The light clusters are assumed to be formed from the nucleons in an equilibrium coalescence. The combination of small overall source size and limited reaction time limit the possible competing species. This is a crucial aspect which has to be considered in the model comparison below.

In the experiments, some very small quantities of $^6$Li and $^7$Li
were found. Note that $^4$Li and $^5$Li are very short-lived unstable
resonances decaying into p+$^3$He or p+$^4$He in $\sim
10^{-24}$~sec. Thus effectively they are included in the other light particle yields in the experiment---if they existed. The upper limit to Li which could be assigned to the source (assuming a three source fitting process for all observed Li) corresponds to an average multiplicity per event of $\sim$0.09. Thus Li occurs in less than 1/10 of the events and cannot account for more than 0.6 mass units which is a fractional mass of $\simeq$2\%. Thus Li is ignored in the derivation of ECs from the models because these are discrete entities and 90\%  of the events will not have a Li present for the other particles to interact with. 
We conclude that the proper comparison with experimental results is to employ $Z \leq 2$ as a constraint in the theoretical models.
Obviously, one could ask what causes the suppression of Li. We suggest that the production of Li is time-constrained in the collision. The formation of Li requires sufficient time to reach equilibrium and the HIC is faster. Such a constraint could potentially be described by a nucleation approach, cf.\ Ref.~\cite{wuenschel2014}.

As will be shown below, the ECs have a dependency on the asymmetry of the system. For the particular source we are investigating, we have identified a total proton fraction of $Y_p^{\rm tot}={Z^{\rm source}}/{A^{\rm source}}=0.41$ as a good representative value. If not noted otherwise, we always use this value in the theoretical models. The same value was also used in the analysis of Ref.~\cite{qin2012}. 

\section{Definition and basic aspects of the EC}
\label{sec_definitions}
In this section we give definitions of quantities used later and review some basic properties of the ECs. Let us start with the number density of an ideal, noninteracting Maxwell-Boltzmann gas of nuclear species $i$,
\begin{equation}
n_i^{\rm id}(T,\mu_i) = g_i \left(\frac{M_i T}{2\pi}\right)^{3/2}\exp\left(\frac1T(\mu_i-M_i)\right)  \; .\label{eq_nid}
\end{equation}
The spin degeneracy factor is denoted by
$g_i$, $\mu_i$ is the chemical potential of the
corresponding nucleus including the rest mass $M_{i}$, and $T$ is the temperature.
We will use the ECs of an ideal gas as a reference, given by
\begin{equation}
 K_c^{\rm id}[i]=\frac{n^{\rm id}_i}{{(n_n^{\rm id})}^{N_i}{(n_p^{\rm id})}^{Z_i}} \; .
\end{equation}
If we assume that chemical equilibrium (i.e., NSE) is established,
\begin{equation}
 \mu_i= N_i \mu_n + Z_i \mu_p \;, \label{eq_nse}
\end{equation}
one finds that the contribution of the chemical potentials to $K_c^{\rm id}$ cancels out, and that it is thus only a function of temperature,
\begin{equation}
 K_c^{\rm id}[i]=K_c^{\rm id}[i](T) \; .
\end{equation}
Therefore it is also composition independent, i.e., it does not depend on which other particles besides $n$, $p$, and $i$ are included as degrees of freedom. 

For example for the $\alpha$ particle one has
\begin{equation}
K_c^{\rm id}[\alpha] = \frac{n_\alpha^{\rm id}}{{(n_n^{\rm id})}^2 {(n_p^{\rm id})}^2} \; . \label{eq_kcidalpha}
\end{equation}
By using Eq.~(\ref{eq_nid}) in Eq.~(\ref{eq_kcidalpha}) one obtains 
\begin{equation}
K_c^{\rm id}[\alpha] = \frac1{2^4} \left(\frac{2\pi}{T}\right)^{9/2}
\left(\frac{M_\alpha}{M_n^2M_p^2}\right)^{3/2}
\exp{\left(\frac{B_\alpha}{T}\right)}\; ,\label{eq_kcidalpha_expl}
\end{equation}
where $B_\alpha$ is the total binding energy of the $\alpha$ particle.
For arbitrary species $i$ we have
 $B_i=N_i M_n + Z_i M_p - M_i$.  
For the values of the binding energies, respectively masses, in the ideal gas case we use Ref.~\cite{AudiWapstra}. 

We want to emphasize that, as long as equilibrium is maintained in the system, all deviations of $K_c$ from $K_c^{\rm id}$ are just due to deviations from classical ideal gas behavior. In turn, as soon as one has deviations from the classical ideal gas behavior, either due to interactions, or due to Fermi-Dirac (or Bose-Einstein) statistics, there will be a remaining dependency on the chemical potentials $\mu_n$ and $\mu_p$. Below, instead of the chemical potentials, we will use the baryon number density $n_B=n_n+n_p+\sum_i n_i A_i$, with $A_i=Z_i+N_i$, and the total proton fraction $Y_p^{\rm tot}=(n_p+\sum_i n_i Z_i)/{n_B}$ as state variables, where the sum over $i$ denotes all considered nuclei. Even in the ideal case, the relation between $(\mu_n,\mu_p)$ and $(n_B,Y_p^{\rm tot})$ depends on which nuclei are included.
Nevertheless, in the ideal, classical case the ECs are still just a function of temperature, even if they are presented as a function of density by using Fig.~\ref{fig_tnb}. Conversely, in the nonideal case, the ECs will depend on the composition and density of the system, in addition to the temperature dependence. 

\section{Dependencies of ECs relevant for the model comparison}
\label{sec_depend}
Before comparing different SN EOS models, we want to identify the dependency of ECs on selected aspects of the EOS which are relevant for this comparison. We are especially interested in aspects which were not discussed in Ref.~\cite{qin2012}. For this analysis we use the model of HS \cite{hempel10} as a starting point and then employ several modifications to it. Obviously, the specifics of this dependence could be different in other models. However, in the present section we just want to identify and illustrate all aspects which can have an influence on the ECs.

\subsection{Brief description of the HS EOS}
\label{sec_hs}
First we give a brief summary of the HS EOS, originally formulated in Ref.~\cite{hempel10}. The HS EOS describes SN matter as a chemical mixture of nuclei and unbound nucleons in NSE. Nuclei are described as classical Maxwell-Boltzmann particles, nucleons with a relativistic mean-field (RMF) model employing Fermi-Dirac statistics. Several thousands of nuclei are considered, including light ones. Their binding energies are either taken from experimental measurements \cite{AudiWapstra} or from various theoretical nuclear structure calculations \cite{Lala99,Moller95}. The following medium modifications are incorporated for nuclei: screening of the Coulomb energies by the surrounding gas of electrons, excited states in the form of an internal partition function, and excluded volume effects. 

\begin{figure}
\begin{center}
\includegraphics[width=0.825\columnwidth]{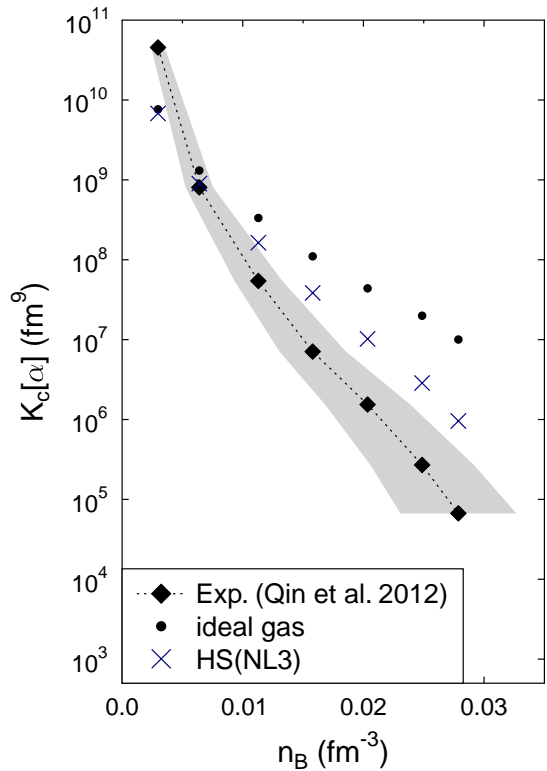}
\caption{\label{fig_kalpha_nl3}(Color online) EC for the $\alpha$ particle from experiments (black diamonds), 
an ideal gas EOS (black dots), and for the HS EOS, applying the NL3 interactions \cite{Lala97} (dark blue crosses). The grey band is the experimental uncertainty of the density determination.}
\end{center}
\end{figure}
The black diamonds in Fig.~\ref{fig_kalpha_nl3} show the equilibrium
constant for the $\alpha$ particle determined from HIC experiments as
published in Fig.~3 of Ref.~\cite{qin2012}. The seven data points
correspond to the averaged experimental data set, and are presented 
as a function of the extracted baryon number density
in a very similar way as in Ref.~\cite{qin2012}. The black dots in 
Fig.~\ref{fig_kalpha_nl3} show $K^{\rm id}_c[\alpha]$ of
Eq.~(\ref{eq_kcidalpha}), i.e., the same quantity but obtained from a
noninteracting ideal gas NSE model using the same temperatures as in
Ref.~\cite{qin2012}. Note again that $K^{\rm id}_c[\alpha]$ is 
independent of the considered particle degrees of freedom and 
only a function of temperature; see Eq.~(\ref{eq_kcidalpha_expl}). Thus in
Fig.~\ref{fig_kalpha_nl3}, the dependency of $K^{\rm id}_c[\alpha]$
on density actually has to be seen as the dependency on the
corresponding temperatures due to the tight correlation between temperature and density
in the experimental data, as shown in Fig.~\ref{fig_tnb}. 

As a starting point, we compare our results for the HS EOS
\cite{hempel10} with the experimental results for the equilibrium
  constant of the $\alpha$ particle given in Ref.~\cite{qin2012}. In Ref.~\cite{qin2012}, the HS model with the RMF interactions NL3 \cite{Lala97} was selected. The ECs for this model are given by the dark blue crosses in Fig.~\ref{fig_kalpha_nl3}. They are calculated from the particle densities, which were taken from the publicly available electronic data tables of the HS(NL3) EOS.\footnote{See \texttt{http://phys-merger.physik.unibas.ch/\midtilde hempel/eos.html}\label{eospage}.} The results are different from those published in Ref.~\cite{qin2012} for the same model. The source of this deviation was a misinterpretation of definitions in the HS EOS in Ref.~\cite{qin2012}. An additional deviation came from the method used to interpolate the tabulated EOS data.
It is important to point out that in Fig.~\ref{fig_kalpha_nl3} the HS(NL3) model converges to the ideal gas limit at low densities and temperatures. The results presented for this model in Ref.~\cite{qin2012} for HS were not correct.

\subsection{Dependence on asymmetry}
In this section, we start to identify factors which influence the
ECs, by investigating the role of the asymmetry,
using the HS EOS. Above we presented updated results for the NL3
interaction, because this was also chosen in
Ref.~\cite{qin2012}. However, it is known that this RMF
  parametrization is in disagreement with some constraints for the
saturation properties of bulk nuclear matter; see, e.g.,
Ref.~\cite{fischer14}. In particular, it has an unrealistically
  stiff density dependence of the symmetry energy. Therefore in the following, for the HS model we will use the RMF interaction DD2 \cite{typel09} instead. This interaction gives a very satisfactory agreement with many other experimental constraints \cite{fischer14}. The model with nuclei is called the HS(DD2) EOS. Note that full SN EOS tables are available for HS(DD2), and have already been employed in SN simulations \cite{fischer14}. 

\begin{figure}
\begin{center}
\includegraphics[width=0.825\columnwidth]{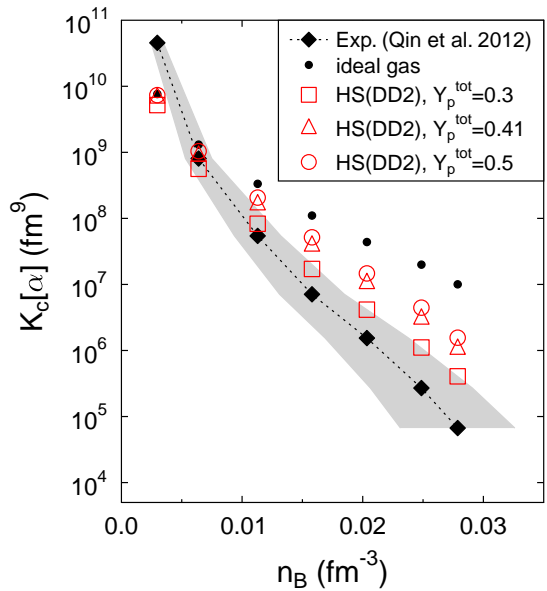}
\caption{\label{fig_kalpha_1}(Color online) EC for the $\alpha$ particle from the HS EOS, applying the DD2 interactions \cite{typel09} (red open symbols). The red circles and squares show variations of the total proton fraction from the value 0.41 used in all other calculations.}
\end{center}
\end{figure}
In Fig.~\ref{fig_kalpha_1}, the EC
  $K_{c}[\alpha]$ is depicted as a function of the baryon density
  $n_{B}$. The red symbols show the
theoretical results for different values of the total proton
fraction from HS(DD2). The red triangles are for $Y_p^{\rm tot}=0.41$, the red
squares for $Y_p^{\rm tot}=0.3$ and the red circles for $Y_p^{\rm
  tot}=0.5$. For $Y_p^{\rm tot}=0.41$ (the standard case) the results
are similar to HS(NL3) presented in Fig.~\ref{fig_kalpha_nl3}. 
The asymmetry, denoted by the total proton fraction $Y_p^{\rm tot}$ of
the system, has an effect on the EC. A
higher $Y_p^{\rm tot}$ increases $K_c[\alpha]$.
 This has to be seen as
an indirect effect: a change of the total proton fraction induces changes of the partial densities of all particles. But as was shown in Sec.~\ref{sec_definitions}, the ECs have no density or asymmetry dependence in the ideal gas case, because the changes of the particle densities cancel out. In contrast, in an interacting model, there is also a change of the strength of the interactions, leading to modifications of $K_c$.
Due to the dependency observed here, the choice of $Y_p^{\rm tot}$ is important. For the remainder of the article we are using $Y_p^{\rm tot}=0.41$, corresponding to the value extracted from the 
experiment; see Sec.~\ref{sec_experiment}.

\subsection{Role of Coulomb effects}
\begin{figure}
\begin{center}
\includegraphics[width=0.825\columnwidth]{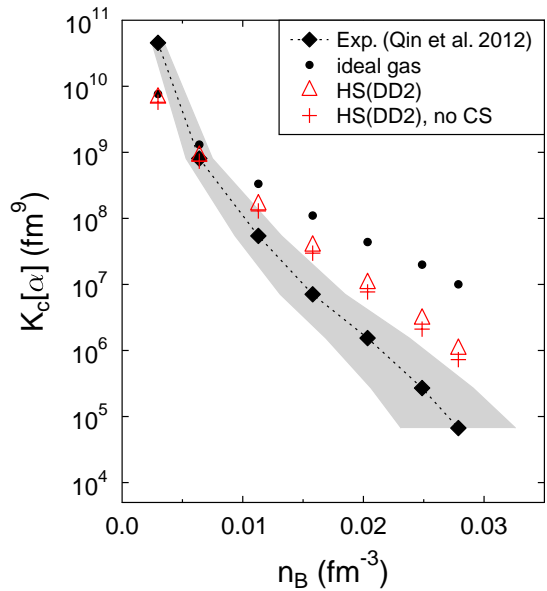}
\caption{\label{fig_kalpha_2}(Color online) EC for the $\alpha$ particle from the HS(DD2) EOS with (red triangles) and without Coulomb screening (CS, red pluses).}
\end{center}
\end{figure}
In SN matter, the Coulomb interactions of nuclei are screened by the
surrounding electrons. This effect is taken into account in some EOS
models, and also in the HS EOS. The Coulomb screening is an
interaction which favors the formation of nuclei with high charge
  as compared to the no-screening case. Thus it also leads to an enhancement of $\alpha$ particles compared to neutrons and protons. In HICs, there is no background of electrons, and thus there should not be any Coulomb screening. 
In addition to the strong interaction between the different constituents, actually there is also Coulomb repulsion which, in principle, contributes to the energy of charged species.
 However, to take this into account is beyond the scope of the present investigation. 

The case of the HS EOS without Coulomb screening is shown in 
Fig.~\ref{fig_kalpha_2} by the red pluses. Switching off the
Coulomb screening leads to a reduction of $K_c[\alpha]$ and the
agreement with the experimental data gets marginally better. However, the
  effect is very small due to the low charge of the $\alpha$ particle. Because it is more realistic, in all following cases of the HS model to be discussed, the Coulomb screening is switched off. Also in all other models considered, we will check the implementation of Coulomb interactions.

\subsection{Particle degrees of freedom}
\label{sec_dofs}
Next we investigate the effect of the included particle degrees of
freedom. Note again, that there would be no composition dependence in
the ideal case. All composition effects on $K_c$ in
nonideal systems are only indirect. A change of the number of
  particle species as relevant degrees of freedoms can change partial densities and thus leads to a change of the strength and form of the interactions, which influences $K_c$. 
In general, all clusters that are allowed to be formed from the source in the experiment should be included in the model calculation. The simplest and most obvious constraint is that the nuclei formed cannot have a mass above the sum of the two colliding nuclei. 
However, as we have argued in Sec.~\ref{sec_experiment}, for comparison to the data which we are considering, the composition should be much more constrained, namely to $Z\leq 2$ 

\begin{figure}
\begin{center}
\includegraphics[width=0.825\columnwidth]{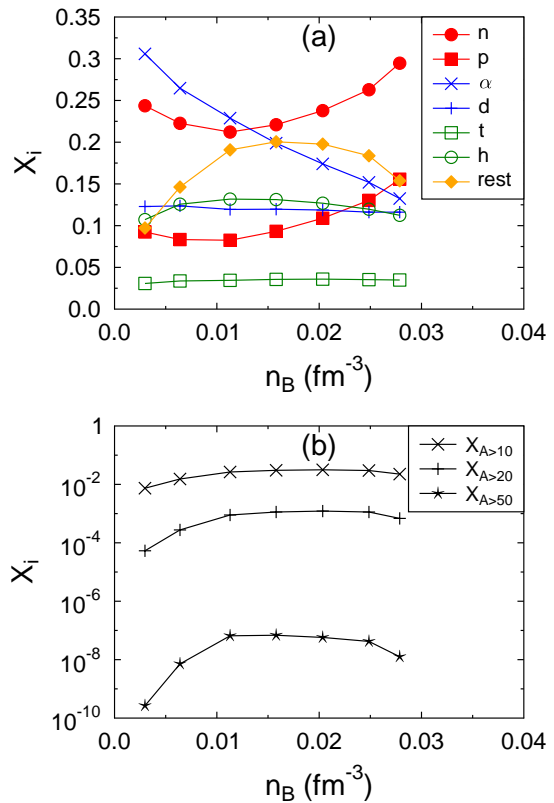}
\caption{\label{fig_compo}(Color online) Composition in the HS(DD2) EOS without Coulomb screening. (a) Mass fractions of neutrons, protons, $\alpha$ particles, deuterons, tritons, helions and the sum of the rest, i.e., all other nuclei on a linear scale. (b) Summed mass fractions of nuclei above mass number $A=10$, 20, and 50 on a logarithmic scale.}
\end{center}
\end{figure}
\begin{figure}
\begin{center}
\includegraphics[width=0.825\columnwidth]{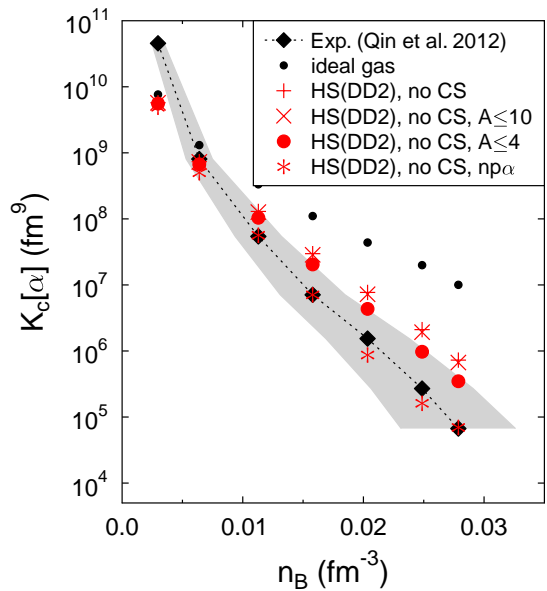}
\caption{\label{fig_kalpha_3}(Color online) EC for the $\alpha$ particle for different compositions of the HS(DD2) EOS, always without Coulomb screening (CS).}
\end{center}
\end{figure}
\begin{figure*}
\begin{center}
\includegraphics[width=1.7\columnwidth]{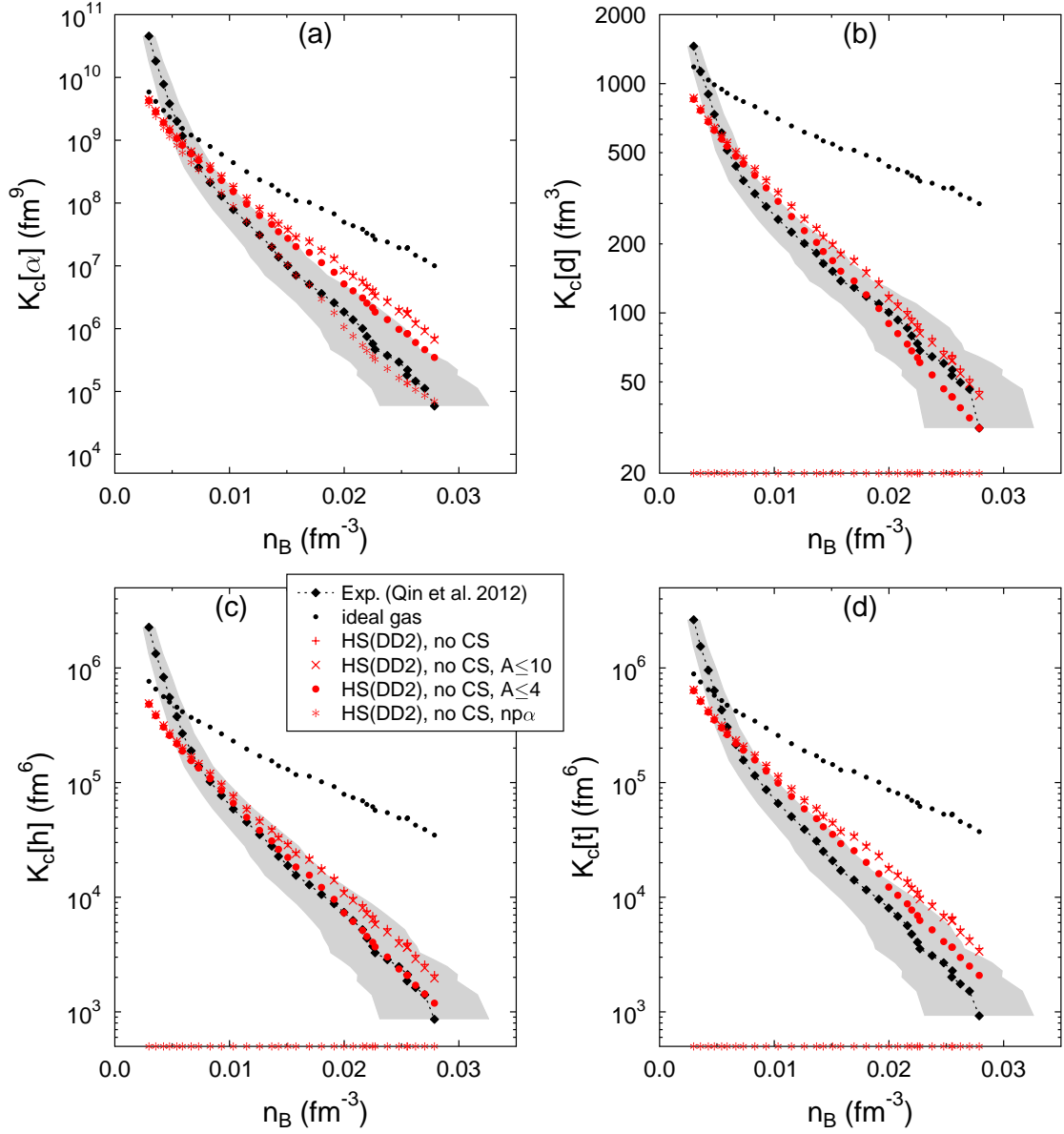}
\caption{\label{fig_kall_hs}(Color online) ECs for (a) $\alpha$ particles, (b) deuterons, (c) helions, and (d) tritons from the full experimental data set (black diamonds) and for different compositions of the HS(DD2) EOS, always without Coulomb screening (CS). The ECs of nuclei which are not included are put on the $x$ axis.}
\end{center}
\end{figure*}
Before discussing the composition dependence of the equilibrium
constant, we will have a look at the chemical composition itself
which is shown in Fig.~\ref{fig_compo} as a function of the
  density. In the HS(DD2) model, there is a small contribution of
very heavy nuclei, as can be seen in the lower panel of Fig.~\ref{fig_compo}. The top panel of this figure shows the mass fractions $X_i=n_iA_i/n_B$ of the most important light nuclei with $Z\leq 2$, nucleons, and the summed mass fraction of all other nuclei (rest). The bottom panel gives the summed mass fractions of nuclei above mass number $A=10$, 20, and 50. The summed mass fraction of nuclei with $Z>2$ (the ``rest'') can exceed 20\%. It is made up from other light and intermediate nuclei, with an exponential distribution as a function of mass number. Helium and lithium isotopes give the largest contributions.
The mass fraction of nuclei with $A>10$ still can exceed 3\% for some of the conditions. The mass fraction of nuclei with $A>20$ is only around or below $10^{-3}$; see Fig.~\ref{fig_compo}. Nuclei with $A>50$ are highly suppressed, and their abundance is completely negligible. 

Now we return to the EC of the $\alpha$ particle.
The red crosses in Fig.~\ref{fig_kalpha_3} show $K_c[\alpha]$ if
only nuclei with $A\leq10$ are considered. The contribution of nuclei
with $A>10$ is too small to have a strong effect on $K_c[\alpha]$. For the filled red circles in Fig.~\ref{fig_kalpha_3}, only  nuclei with $A\leq4$ and $Z\leq2$ are considered: neutrons, protons, deuterons, helions, tritons, and $\alpha$ particles. This case is motivated by the arguments given in Sec.~\ref{sec_experiment} that the formation of all heavier nuclei is suppressed in the experiment.
We find that this limited composition induces some small, but notable differences compared to the case of $A\leq10$ (red crosses) or the unconstrained composition (red pluses). For the HS(DD2) model it leads to better agreement with the experiment.
For the red asterisks (denoted by ``np$\alpha$'') only neutrons, protons, and $\alpha$ particles are included. It is clear that this limited composition is only considered for illustrative purposes. Other nuclei, e.g., the deuteron, are in fact seen abundantly in the experiment. We include this case here, because the limited np$\alpha$ composition (plus a representative heavy nucleus) is used, e.g., in the STOS and the LS EOS. The results shown in Fig.~\ref{fig_kalpha_3} for this case demonstrate once more that the considered degrees of freedom can have a big effect on $K_c$. 
Even though the agreement of the np$\alpha$ case of the HS(DD2) EOS
with the experiment is excellent, we do not think this model should be
regarded as more realistic. Instead this agreement rather has to be
seen as a coincidence. Obviously, the np$\alpha$ EOS would fail to
explain the nonzero values of $K_c[d]$, $K_c[t]$, or $K_c[h]$.

This is illustrated in Fig.~\ref{fig_kall_hs}. In contrast to the figures above, we are comparing with the full experimental data set, which also includes the ECs of deuterons, helions and tritons. We are considering again the four different assumptions for the composition as in Fig.~\ref{fig_kalpha_3}. Strictly speaking, the np$\alpha$ case makes the prediction that $K_c[d]$, $K_c[h]$, and $K_c[t]$ are all zero. Therefore we have put the corresponding symbols on the $x$ axis. If instead the right corrections for the comparison to HIC experimental data ($A\leq 4$, no Coulomb screening) are applied, the HS EOS model is in quite good agreement with the experimental data. The most apparent deviation appears for the $\alpha$ particle, where the HS model predicts slightly larger values, but still mostly within the experimental error bars. 

\section{Description and modifications of other EOS models}
\label{sec_other}
In a similar way as for the HS EOS, in the following we give a brief description of various other SN EOS models, and modify them such that they can be applied for the HIC experiments. We will not go through all the steps which were done in Sec.~\ref{sec_depend}, but concentrate on the peculiarities of each model.

\subsection{STOS}
As in the HS EOS, the STOS EOS of H.~Shen \textit{et al}.\ \cite{shen98,stos} also
uses a RMF interaction for the nucleons.
However, only one EOS table is available which is based on the
TM1 parametrization \cite{Suga94}. The STOS EOS employs neutrons,
protons, and $\alpha$ particles as explicit particle degrees of
freedom. Nucleons are described by Fermi-Dirac statistics, alpha
particles with Maxwell-Boltzmann statistics and excluded volume
corrections. Excited states of $\alpha$ particles are not considered. The
  formation of heavy nuclei is treated in the single-nucleus
  approximation (SNA). The properties of the representative nucleus are obtained from Wigner-Seitz cell calculations within the Thomas-Fermi approximation for parametrized density distributions of nucleons and $\alpha$ particles. 
Its translational energy is not taken into account.

\begin{figure}
\begin{center}
\includegraphics[width=0.825\columnwidth]{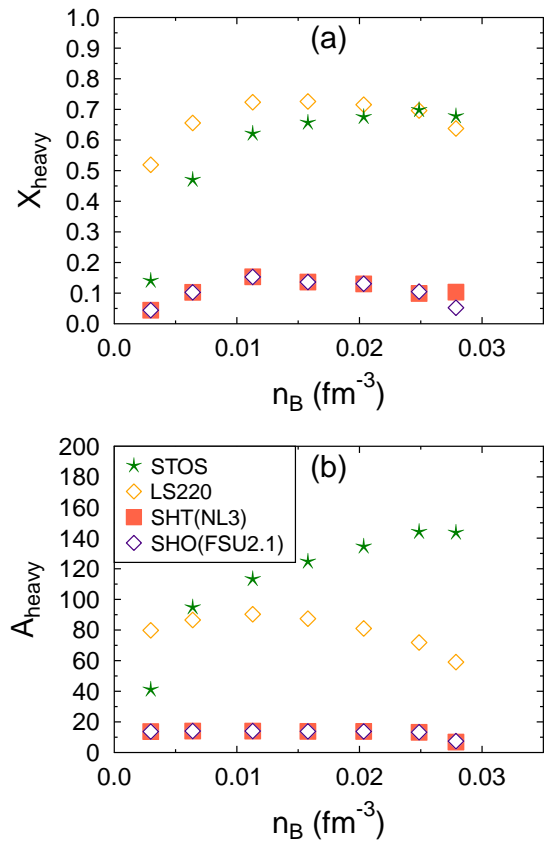}
\caption{\label{fig_compo_all}(Color online) (a) Mass fraction of heavy nuclei of the STOS, LS220, SHT(NL3), and SHO(FSU2.1) EOSs, for the conditions extracted from the experiment. (b) Corresponding average, respectively representative, mass number of heavy nuclei.}
\end{center}
\end{figure}
In Fig.~\ref{fig_compo_all}, we show the mass fraction of heavy nuclei of the STOS EOS for the thermodynamic conditions corresponding to Fig.~\ref{fig_tnb}. For most of the data points, about 50\% or more of all nucleons are bound in heavy nuclei. These heavy nuclei have masses in the range from 40 to 140 which is shown in the bottom panel.
In the unmodified HS model, such heavy nuclei are not found, but light and intermediate nuclei are favored instead; see Fig.~\ref{fig_compo}. Heavy nuclei also appear in HS, but only at higher densities. The appearance of heavy nuclei in STOS can thus be interpreted as a compensation effect for missing light nuclei, but might also depend on different descriptions of temperature effects in heavy nuclei (compare also with Ref.~\cite{buyuk13}).

The composition found for STOS cannot be produced in the experiment. Therefore we have repeated the calculations of STOS, 
suppressing 
the formation of heavy nuclei by assuming uniform nucleon and $\alpha$ particle distributions. In this situation, one obtains the np$\alpha$ composition which was also considered as one of the HS modifications in Sec.~\ref{sec_dofs}. We think that the np$\alpha$ case represents the appropriate limit of the STOS EOS for the HIC we are investigating, because other light nuclei are not considered in the model of STOS.

The np$\alpha$ model of STOS is slightly different than that of HS. Different RMF interactions are employed in the two models for the nucleons, the excluded volume prescriptions are different and no Coulomb screening is used in STOS for the $\alpha$ particles. The last point means that we do not have to apply any further corrections for the Coulomb energies, as done in the HS model.
We have checked that our calculations of the np$\alpha$ case of the STOS model agree with the published data in regimes where the full STOS EOS gives the np$\alpha$ composition by itself. The agreement was typically up to the third digit in mass fractions and second digit for ECs.

\begin{figure}
\begin{center}
\includegraphics[width=0.825\columnwidth]{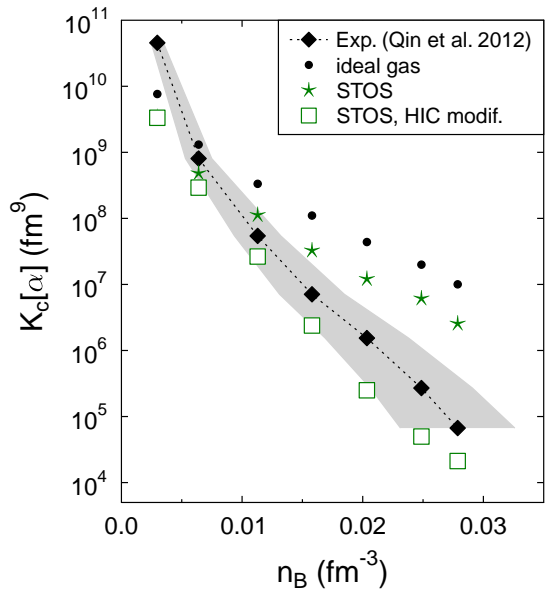}
\caption{\label{fig_kalpha_shen}(Color online) EC for the $\alpha$ particle from the original STOS EOS \cite{shen98} (green stars) and for a modification of it for the conditions in the HIC experiment (green squares), where heavy nuclei are suppressed.}
\end{center}
\end{figure}
In Fig.~\ref{fig_kalpha_shen}, we show the $\alpha$ particle EC, for the 
original STOS model and the STOS np$\alpha$ model, 
 denoted as ``STOS, HIC modif.'' in which we suppressed heavy nuclei.
The model modified for HICs shows a better agreement with the
  experimental data than the uncorrected one. The results are close or within the error bars. Independent of whether or not these corrections are applied, the STOS EOS does not include any other light nuclei than $\alpha$ particles, and thus cannot explain the other ECs.
  
The EOS of Furusawa \textit{et al}.\ \cite{furusawa11, furusawa13} represents an extension of the STOS EOS where other light nuclei and a distribution of heavy nuclei are already included. Furthermore, it incorporates both Pauli-blocking shifts and excluded volume effects to dissolve light clusters at high densities \cite{furusawa13}. This model has already been used to explore the effect of light nuclei in core-collapse SN simulations \cite{furusawa13b}. We expect that this EOS gives a better agreement with the experimental data. A tabulated version of this EOS is not yet publicly available, therefore here we are using only the STOS EOS.

\subsection{LS}
The EOS of Lattimer and Swesty \cite{lattimer91} exists for three different (nonrelativistic) parametrizations of the nucleon interaction, which are usually denoted according to their value of the incompressibility $K$ of 180, 220, and 375~MeV. The last value is now considered as incompatible with experimental data such as those extracted from measurements of the giant monopole resonance. The EOS with $K=180$~MeV leads to a too low maximum mass of neutron stars, which is not compatible with latest astrophysical observations. Thus we only investigate the version with $K$=220~MeV, which we call LS220.

The LS EOS considers nucleons, $\alpha$ particles and heavy nuclei in SNA as
degrees of freedom. The latter are described with a liquid-drop
model. For the nucleons, nonrelativistic Fermi-Dirac statistics are
used, for the $\alpha$ particle Maxwell-Boltzmann statistics and excluded
volume corrections. Excited states of $\alpha$ particles are not considered. Heavy
nuclei are also described by Maxwell-Boltzmann statistics,
however, with fixed mass number in the translational energy.

The composition of matter in the LS220 model is shown in Fig.~\ref{fig_compo_all} by the open orange diamonds. One has a similar situation to that of the STOS EOS: there is a huge fraction of very heavy nuclei, which cannot be produced in the HIC experiments we are considering. Thus in a similar way to which we modified the STOS EOS, we have modified the LS calculations to suppress heavy nuclei. As before, this means one is left with the np$\alpha$ model. In this case the LS model does not include any Coulomb screening due to surrounding electrons. 

\begin{figure}
\begin{center}
\includegraphics[width=0.825\columnwidth]{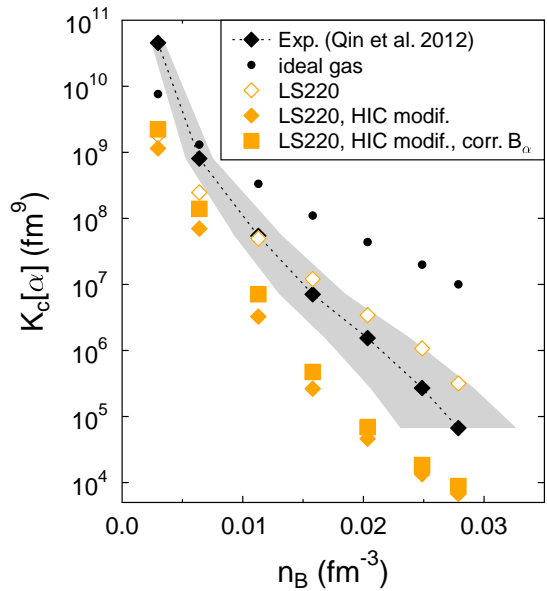}
\caption{\label{fig_kalpha_ls}(Color online) EC for the $\alpha$ particle from the
  original LS220 EOS \cite{lattimer91} (open orange diamonds) and for
  a modification of it for the conditions in the HIC experiment
  (filled orange diamonds), where heavy nuclei are suppressed. For the full orange squares in addition a correction for the binding energy of the $\alpha$ particle was implemented.}
\end{center}
\end{figure}
The two cases are compared with the experimental data in
Fig.~\ref{fig_kalpha_ls}. The unmodified model agrees quite well,
however, this has to be seen as a coincidence. The suppression of
  heavy nuclei,  which cannot be formed in the experiment, 
leads to too low values of $K_{\alpha}$ in the thus modified EOS.

It is known in the literature \cite{swesty2005} that 
the binding energy of the $\alpha$ particle is not
  correct in the original LS model \cite{lattimer91}. A value of
28.3~MeV is employed, which is similar to recent experimental
measurements. However, in LS, all energies are measured with respect
to the neutron mass. Therefore the correct value should be $28.3+2
(M_n-M_p)\simeq 30.886$~MeV. The full squares in
Fig.~\ref{fig_kalpha_ls} show the results of the np$\alpha$ EOS with
the corrected binding energy. It leads to a slight enhancement of
$K_c[\alpha]$ and thus a better agreement with the data.
We also note that the prediction of LS220 in this case is similar to, 
though consistently below, that of STOS.

\subsection{G.\ Shen \textit{et al}.}
\label{sec_gshen}
The EOSs of G.\ Shen \textit{et al}.\ are available for two different RMF interactions, NL3 \cite{gshen11nl3} and FSUgold \cite{gshen11fsu}. Because the FSUgold parametrization lead to a maximum neutron star mass of only 1.7 M$_\odot$, an additional phenomenological pressure contribution was introduced, leading to a sufficiently high maximum mass of 2.1 M$_\odot$. This latter variant is called FSU2.1. Even though this distinction is not relevant for the densities we are interested in, we will use FSU2.1 in the following, which we abbreviate ``SHO(FSU2.1).'' The EOS from Ref.~\cite{gshen11nl3} will be abbreviated by ``SHT(NL3).'' 

\begin{figure}
\begin{center}
\includegraphics[width=0.825\columnwidth]{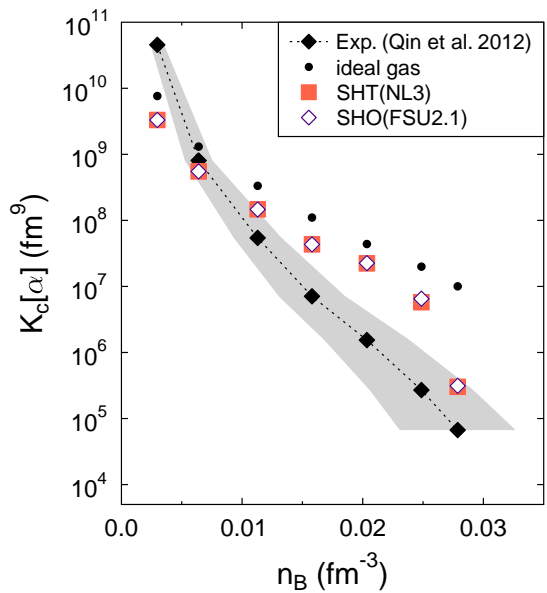}
\caption{\label{fig_kalpha_gshen}(Color online) EC for the $\alpha$ particle from two
  EOSs from G.~Shen, one with NL3 interactions \cite{gshen11nl3}
  (dark orange squares) and one with FSUgold interactions
  \cite{gshen11fsu} (violet diamonds).}
\end{center}
\end{figure}

The EOSs of G.\ Shen \textit{et al}.\ are based on different underlying physical
descriptions in different regimes of density and temperature. At
highest densities, there is uniform nuclear matter consisting of only
nucleons which are described by the corresponding RMF model and
Fermi-Dirac statistics. At intermediate densities, the RMF model is
used within Hartree calculations of nonuniform matter, generating a
representative heavy nucleus and unbound nucleons, but no light
nuclei. At lowest densities, a special form of the virial EOS is used
which includes virial coefficients up to second order among nucleons
and $\alpha$ particles. Note that the virial EOS is not using Fermi-Dirac
statistics for the nucleons, but only incorporates corrections for it
as part of the virial coefficients. Mass 2 and 3 nuclei are not
included as explicit degrees of freedom, but 8980 nuclei with mass
number $A\geq 12$ are included. The contribution of heavy nuclei in NSE is modeled
as a noninteracting Maxwell-Boltzmann gas without considering
  excluded volume effects.
Coulomb screening is included for heavy nuclei, but not for the $\alpha$ particles. Note that $\alpha$ particles are only present in the virial part of the EOS, which is completely independent of the RMF interactions.

The compositions of the two EOSs of G.~Shen are shown in
Fig.~\ref{fig_compo_all} by the dark orange and violet symbols. Except
  for the point at highest densities, the two models SHT(NL3) and
SHO(FSU2.1) give very similar results. This is because they are both in the virial EOS regime, using an identical model description. For the last data point, in both models we have $A_{\rm heavy}<12$ and slightly different results for $X_{\rm heavy}$. This is probably an indication for the onset of the transition to uniform matter or the Hartree calculations. For all conditions, there is a notable contribution of heavy nuclei on the order of 10\% which have $A \lesssim 15$.
The nuclei which were found in the unmodified HS model (see Fig.~\ref{fig_compo}) had lower mass numbers. On the other hand, the $\alpha$ particle fraction is increased in the EOS of G.~Shen \textit{et al}.

The heavy nuclei found in the EOS of G.~Shen \textit{et al}.\ cannot be produced in the experiments. Furthermore, Coulomb screening corrections are applied for their description. Unfortunately it would not be feasible for us to repeat the calculations of G.\ Shen \textit{et al}.\ to remove these heavy nuclei and the Coulomb screening, necessary to reproduce the conditions in HICs. However, note that Coulomb screening is not applied for the $\alpha$ particle. Thus there is at least no direct effect on $K_c[\alpha]$. Even for the HS model,
 where a direct correction from Coulomb screening of the $\alpha$ particle was made,
 the effect was not very strong; see Fig.~\ref{fig_kalpha_2}. We expect it to be less for the G.\ Shen EOS.
Note that the compositions of G.\ Shen EOSs are not dominated by heavy nuclei.
  Their abundances are much less than in the unmodified HS, LS or STOS models. Thus we do not expect that the suppression of Coulomb screening and heavy nuclei would have a big effect and think that it is acceptable to apply the unmodified EOSs of G.\ Shen \textit{et al}.\ for the case of HICs.    

In Fig.~\ref{fig_kalpha_gshen}, we show the $\alpha$ particle equilibrium
constants for the two {G.~Shen EOSs. As for the compositions, we
also find for $K_c[\alpha]$ that they are very similar. In general,
apart from the two points at lowest densities, both models give
values for $K_c[\alpha]$ which are too high. 
The results of the G.~Shen
EOSs are rather close to the ideal gas values, and follow a
similar trend. 
The differences from the ideal gas behavior are 
  generated by the experimentally derived virial coefficients.
At highest density,  a sudden decrease of
$K_c[\alpha]$ occurs, which is probably due to the onset of the transition
  from the ``virial regime'' to the intermediate density regime
  without $\alpha$ particles. 
Note that the tabulated EOSs of G.~Shen \textit{et al}.\ are based on a smoothing and interpolation 
procedure \cite{gshen11nl3,gshen11fsu}, which could explain why the transition between the two regimes is smoothed out and not abrupt.

\subsection{gRDF}
\begin{figure*}
\begin{center}
\includegraphics[width=1.7\columnwidth]{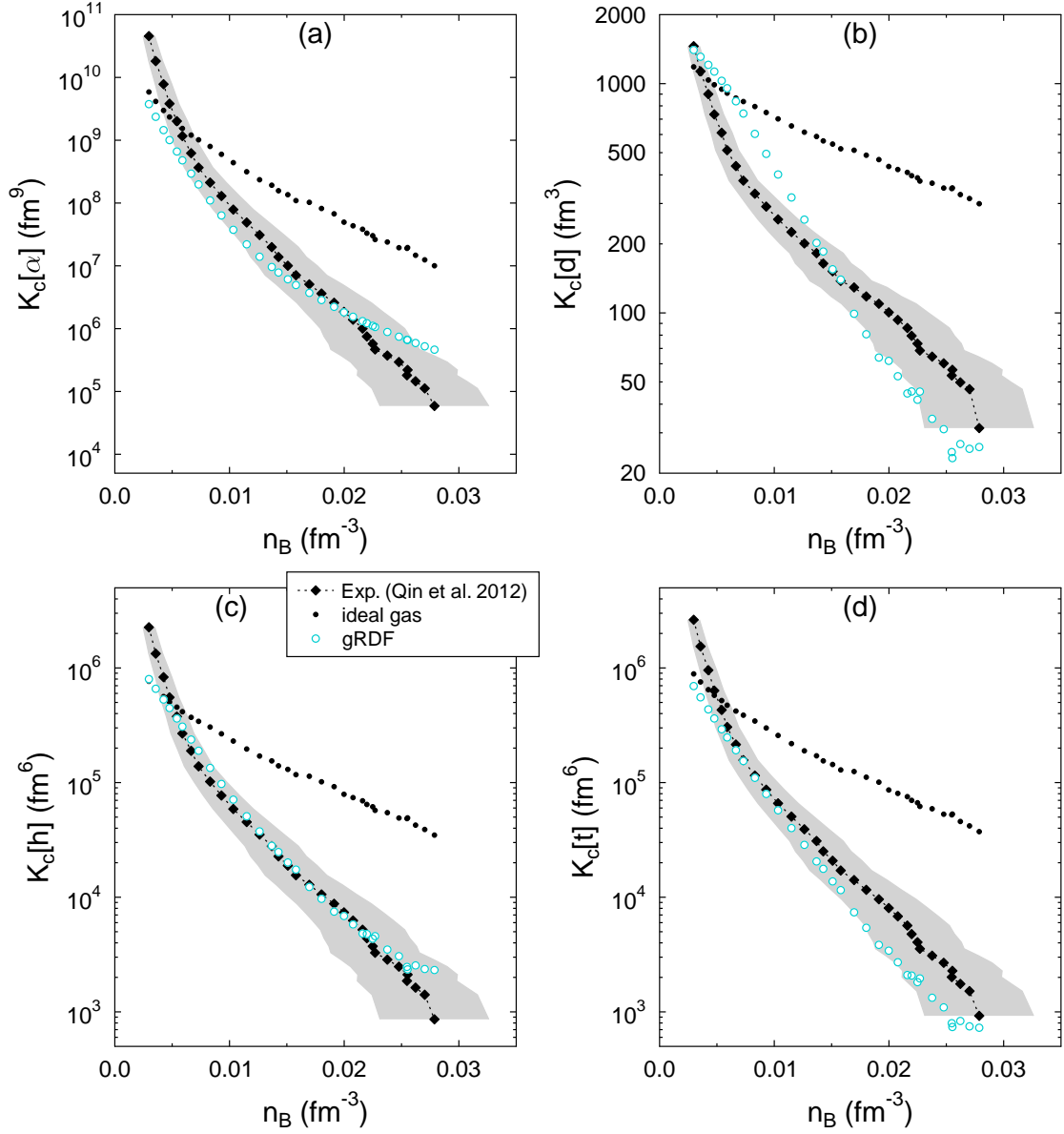}
\caption{\label{fig_k_grmf}(Color online) EC for (a) $\alpha$ particles, (b) deuterons, (c) helions, and (d) tritons from the gRDF model (cyan circles).}
\end{center}
\end{figure*}

The generalized relativistic-density functional (gRDF) was originally
developed in Ref.~\cite{typel09} and further refined in
Ref.~\cite{voskresenskaya2012}. In this model, nucleons, light nuclei, and
heavy nuclei are used as particle degrees of freedom, and all of them
are interacting with each other in a meson-exchange based effective
relativistic mean-field approach. For the nucleons, Fermi-Dirac
statistics are used, for the nuclei Maxwell-Boltzmann. Nucleon-nucleon
scattering correlations are included as explicit degrees of freedom
and represented by temperature dependent effective resonances in
  the nucleon-nucleon scattering continuum. In the calculation of the ECs their contributions are counted as free protons and neutrons, because the nucleons in these states are unbound. 

In addition, medium dependent binding energy shifts of nuclei are
incorporated. These are either extracted from quantum statistical
calculations of R\"opke \cite{typel09} for light nuclei, where
further details will be given in Sec.~\ref{sec_qs}, or from
Thomas-Fermi calculations for heavy nuclei, not considered here. 
In comparison to Ref.~\cite{typel09}, here we use an updated parametrization of the binding energy shifts of light nuclei.
Details of the new parametrizations are given in Appendix \ref{app_grdf}.
In the calculations presented in the present work, Coulomb shifts and
excited states are omitted and only neutrons, protons, deuterons,
tritons, helions, and $\alpha$ particles are considered as particle degrees of
freedom. 
 
Figure~\ref{fig_k_grmf} shows the results for the ECs
in the gRDF model for the conditions as discussed in Secs.~\ref{sec_experiment} and \ref{sec_depend}.
The results are very sensitive to the choice of the density dependence
of the mass shifts. For the densities and temperatures of the 
experiment, the light clusters, in particular the deuterons, are unbound in the gRDF model and thus the choice of the mass shifts for the unbound clusters is a delicate problem. 
This will be further illustrated below in Sec.~\ref{sec_qs}.

With the present 
energy-density functional, the gRDF model gives very satisfactory results.
The overall behavior of all four ECs from the experiment is reproduced very well, mostly within the error bars. 
At lowest densities there is the discrepancy
which is seen in all models, and which will be discussed
further in Sec.~\ref{sec_summary}. Only for the deuterons some slightly 
stronger deviations are found. Below $n_B\simeq 0.017$~fm$^3$, 
there is overprediction, above underprediction.}

However, for the deuteron there is excellent agreement for the
point at lowest density, in contrast to the results of the models discussed above.
Interestingly, this point is above the NSE value, which we want to comment 
on briefly. The scattering state in the deuteron channel,
which is represented by an effective resonance, gives a negative contribution 
to the total density. If there is an increase of the continuum contribution also
the bound state contribution becomes larger while keeping the sum nearly constant.
Thus the ground state density can be larger in the gRDF model than in the NSE calculation.
Note that this behavior depends on how the shift of the
continuum part is parametrized.

\subsection{QS}
\label{sec_qs}

\begin{figure*}
\begin{center}
\includegraphics[width=1.7\columnwidth]{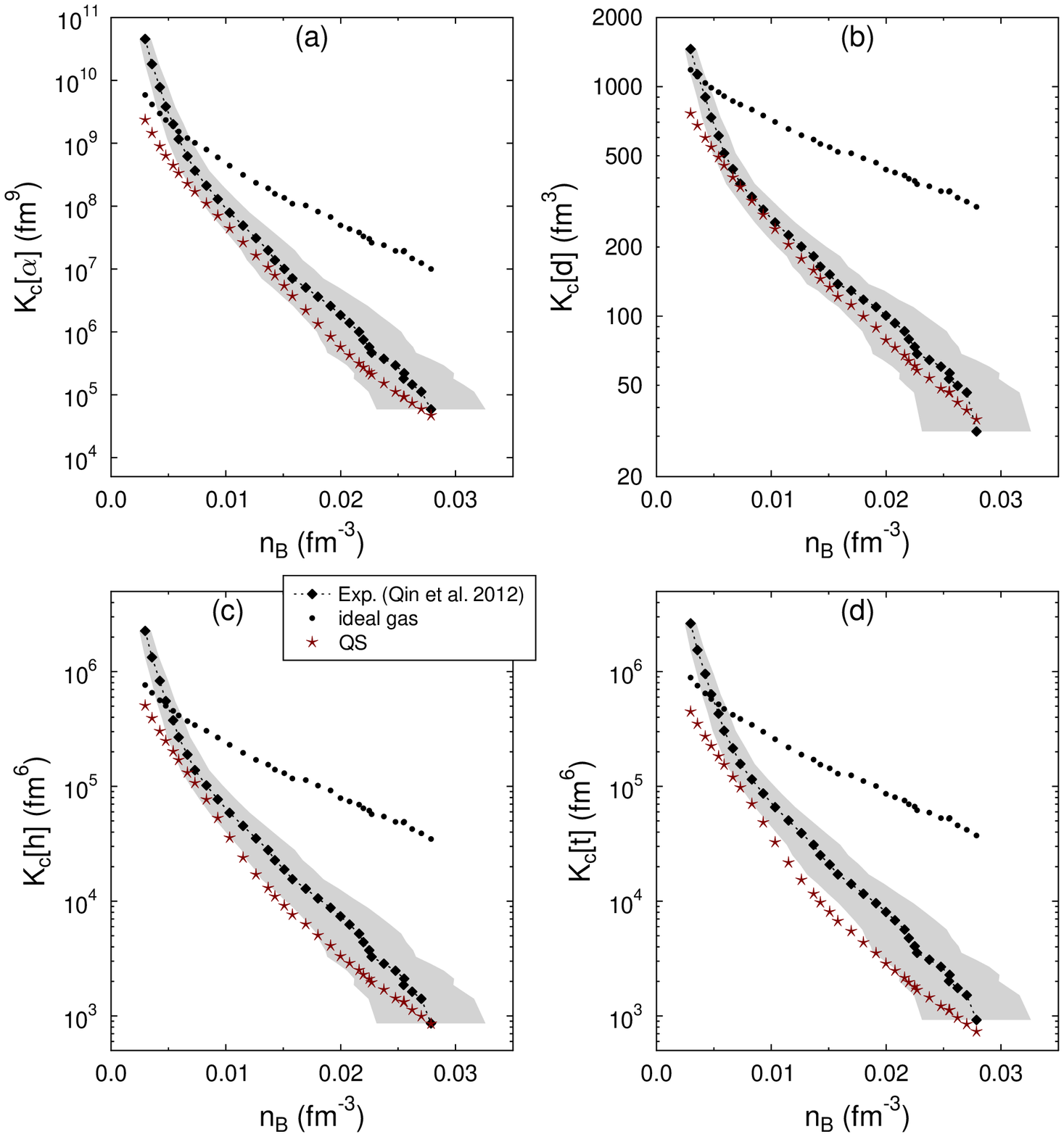}
\caption{\label{fig_k_qs}(Color online) EC for (a) $\alpha$ particles, (b) deuterons, (c) helions, and (d) tritons from the quantum statistical (QS) model (brown stars).}
\end{center}
\end{figure*}

To treat the contributions of cluster interactions in the EOS, instead
of the semiempirical excluded volume concept as used, e.g., in the HS EOS, a 
microscopic treatment can be given, starting from a systematic quantum 
statistical (QS) approach; see Refs.~\cite{cmf2,RMSa,RMSb}. 
Effects of the nuclear medium on the cluster under consideration such as 
self-energy shifts and  Pauli blocking are taken into account, leading,
e.g., to the merging of bound states with the continuum of scattering states at
increasing density (the so-called Mott effect). 

The QS model is described in detail in Refs.~\cite{roepke09,roepke11a}. It is
based on the thermodynamic Green-function method and uses an effective
nucleon-nucleon interaction. Starting from exact expressions for the spectral 
function,
a cluster decomposition of the self-energy can be performed. The total baryon 
density is decomposed into contributions
of different clusters (nuclei) which are obtained from an effective wave 
equation containing medium effects;
see Appendix \ref{app:QS}. In particular, self-energy and Pauli blocking appear 
as leading terms of interaction with the surrounding medium.
This way, single nucleon states and cluster states become quasiparticles, with 
medium dependent energies and medium-modified 
wave functions.

The medium modifications of the clusters can be determined from the in-medium 
Schr\"odinger equation; see Eq.~(\ref{waveA}) in Appendix \ref{app:QS}. 
The binding-energy shift $\Delta E^{\rm qu}_i({\bf P};T,n_B,Y_p^{\rm tot})$ 
of each cluster $i$ (depending on $T$, $n_B$, $Y_p^{\rm tot}$, and center-of-mass (c.o.m.)
momentum $\bf P$) with respect to the energy in the vacuum, 
contains the self-energy shift $\Delta
E^{\rm SE}_i$, the Coulomb shift $\Delta E^{\rm Coulomb}_i$ (in SN matter), and 
the
Pauli blocking shift $\Delta E^{\rm Pauli}_i$. The latter is
relevant for the decrease of the binding energy of nuclei with increasing
density and determines the Mott densities, where the clusters dissolve. 
Approximations for the cluster formation in the medium have been worked out in 
Refs.~\cite{clustervirial,cmf2,RMSa,RMSb}.
Here we use the recent parametrization of the shifts of the in-medium binding 
energies from Ref.~\cite{roepke11a}. 
The nucleon self-energies in the QS model are evaluated with the RMF 
model DD2 \cite{typel09}, 
and the quasiparticle shifts of light clusters $d,t,h,\alpha$ are calculated 
within a variational approach \cite{roepke09,roepke11a}.

An important issue is the inclusion of excited states. This is not a problem at 
low densities for the ideal gas NSE model and the HS approach as long as excited 
bound states (nuclei) are considered. However, the scattering states are 
essential to obtain the correct second virial coefficient.
The QS approach gives the Beth-Uhlenbeck formula in the low density limit, but 
gives also a generalized Beth-Uhlenbeck formula which takes in-medium effects 
into account \cite{SRS}. The virial EOS of G. Shen \textit{et al}.\ 
\cite{gshen11nl3,gshen11fsu} uses also the correct second virial coefficient so 
that the chemical constants are reduced, 
but mean-field effects are not included consistently so that the results for the 
chemical constants remain near the values of the ideal NSE case. An attempt to construct 
an EOS which treats
simultaneously the virial coefficient and the mean-field contributions has been 
performed in Ref. \cite{voskresenskaya2012}. 

We present results which are derived from 
a generalized Beth-Uhlenbeck approach \cite{SRS} and the cluster-virial 
expansion \cite{clustervirial}. Virial contributions are of relevance for the 
deuteron yields when the temperature is large compared with the binding energy. 
Therefore, the values for $K_c[d]$ depend strongly on the 
approximation for the second virial coefficient, i.e. the treatment of the 
continuum correlations. Similar to the bound states,
the medium modification of the continuum contributions depend on the c.o.m.\ 
momentum $\bf P$. In extension of Ref. \cite{roepke2014b} 
where only the medium modifications at ${\bf P}=0$ have been considered, 
expressions for finite $\bf P$ are given in Appendix~\ref{app:QS}.

With respect to HICs, the fate of the continuum states is not obvious. The 
freeze-out approach which we are considering cannot follow the real time 
evolution of correlations in the continuum. Thus the question arises whether the 
continuum states are contributing to the yields of clusters or to the single 
nucleon yields? We separated the mean-field contributions of the continuum 
states and considered only the residual part of the continuum correlations in 
the respective channel $i$, contributing to the cluster yields.

A particular problem is  the effect of cluster formation in the medium so that 
one has to go beyond 
the mean-field approach, for instance considering the cluster mean-field 
approximation \cite{cmf2,clustervirial}. In principle, the systematic inclusion 
of correlations in nuclear matter should include the self-consistent treatment 
of cluster formation in the self-energy as well as in the Pauli blocking term. 
For example, a systematic approach should also be able to describe alpha-particle 
matter. For discussion see Ref.~\cite{roepke2014b} where an approximate approach 
is given. For the present calculation of the composition of nuclear matter, we 
considered a correlated medium as described in Appendix~\ref{app:QS}.

The results of the QS model are presented in Fig.~\ref{fig_k_qs}, whereas the 
following particles were included as degrees of freedom in the calculations: $n$, 
$p$, $d$, $t$, $h$, and $\alpha$. Overall, the QS model shows excellent agreement with 
the experimental data. The calculated QS values for the chemical constants are 
somewhat lower than the experimental ones, but we mention already that the 
results depend on the approximations performed,  in particular the treatment of 
continuum correlations. We want to emphasize again that the QS model is the only 
approach currently available, which is able to predict the suppression of light 
nuclei at high densities on a microscopic basis of a quantum-statistical 
description. Thus the found agreement is quite remarkable.

\begin{figure*}
\begin{center}
\includegraphics[width=1.7\columnwidth]{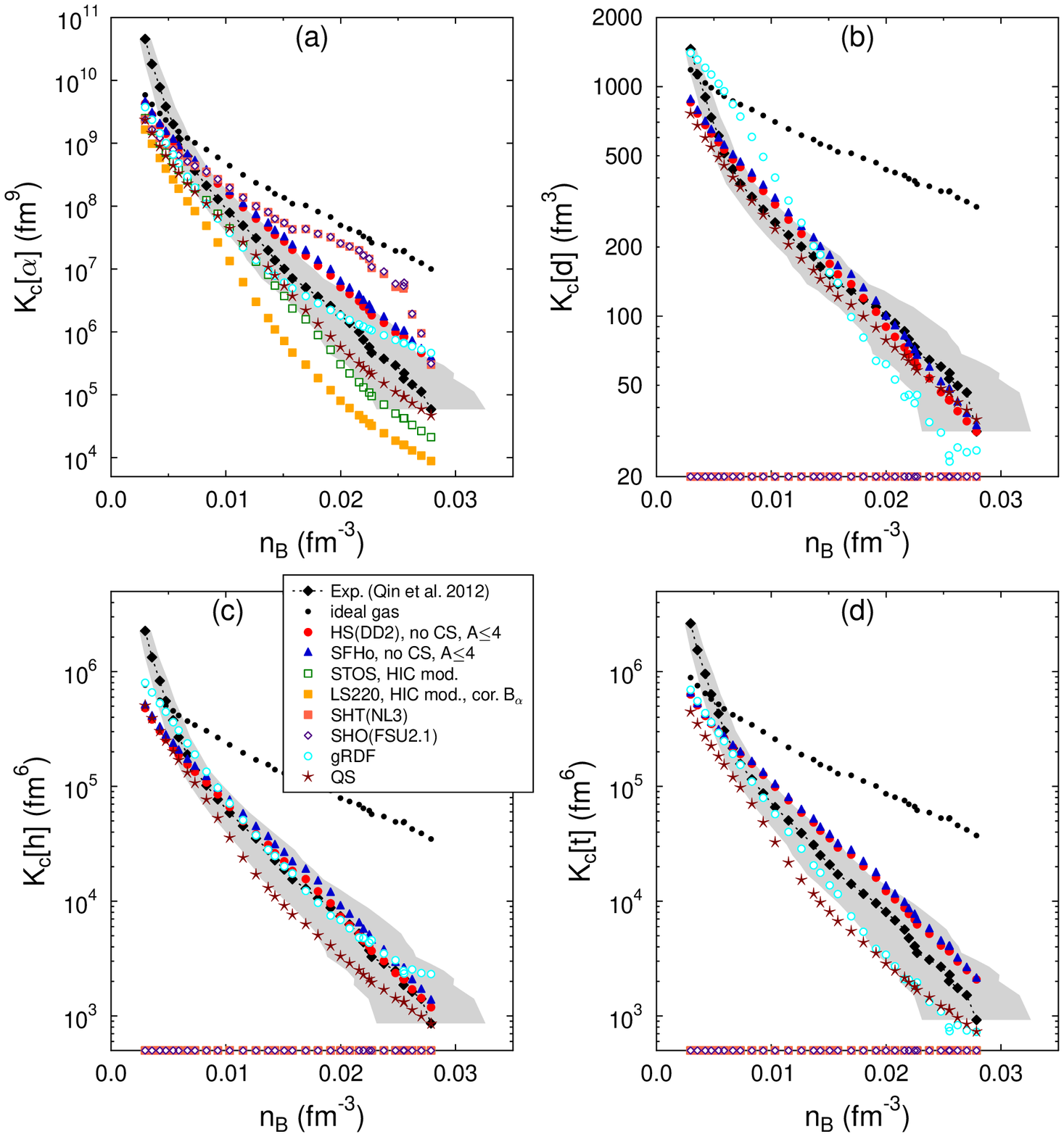}
\caption{\label{fig_kall}(Color online) ECs for (a) $\alpha$ particles, (b) deuterons, (c) helions, and (d) tritons from experiments (black diamonds) in comparison with those of various theoretical models, which are all adapted for the conditions in HIC, as far as possible. The ECs of nuclei which are not included in a model are put on the $x$ axis.}
\end{center}
\end{figure*}
\begin{figure*}
\begin{center}
\includegraphics[width=1.7\columnwidth]{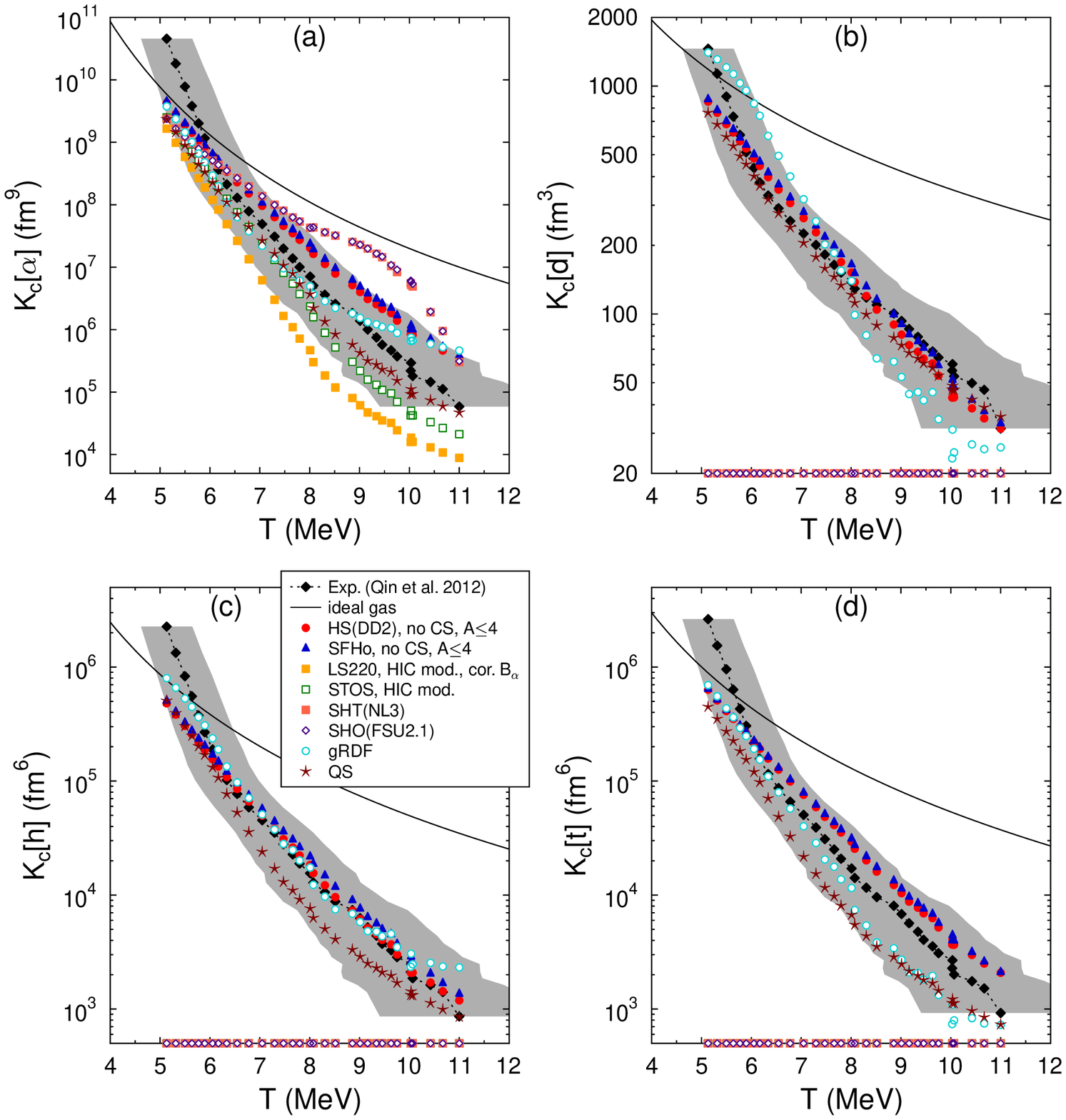}
\caption{\label{fig_kall_t}(Color online) This figure shows the main results of our investigation: EC for (a) $\alpha$ particles, (b) deuterons, (c) helions, and (d) tritons as a function of temperature. The grey band is the experimental uncertainty for the temperature determination. Experimental data (black diamonds) are compared with various different theoretical models, which are all adapted for the conditions in HICs, as far as possible. The ECs of nuclei which are not included in a model are put on the $x$ axis. The black lines show the ECs of the ideal gas model, which are solely a function of temperature.}
\end{center}
\end{figure*}

\section{Constraining SN EOS}
\label{sec_benchmarking}
Figure~\ref{fig_kall} shows the ECs of all the
different models which were introduced above and prepared for the
comparison. The only additional model is
SFHo from Ref.~\cite{steiner13}, which is based on the HS model, but
employs a nucleon-nucleon interaction different from HS(DD2). We
have selected this model as another interesting case from the various
HS models available in the literature \cite{fischer14}, because its
nuclear matter properties are in agreement with various experimental
and theoretical constraints, as are those for the HS(DD2)
EOS. However, this model is constructed such that certain radius
  measurements of neutron stars
were reproduced. It gives more compact neutron stars than the other models considered here. This can be seen as the main characteristic of SFHo. In Fig.~\ref{fig_kall}, the SFHo model leads to very similar results as HS(DD2), and is mostly within the experimental error bars.  

If we compare all the considered EOS, there is an obvious
distinction between two groups of models: in the first group, there
are np$\alpha$ models (LS220 and STOS) and models which do not include
other light nuclei than $\alpha$ particles as explicit degrees of freedom
(SHT(NL3), SHO(FSU2.1)). As before, we have put the data points for
the ECs of nuclei which are not included in these
models onto the $x$ axis.
Strictly speaking, such models assume that the corresponding $K_c[i]$ are zero. In the second group, there are models with $n,p,d,t,h,\alpha$ composition (HS(DD2), SFHo, gRDF, and QS), i.e., models that include all nuclei observed in the experiment. 

We can compare $K_c[\alpha]$ for all the models.
 If we consider the overall range of the theoretical predictions we observe a convergence at low densities, approaching the ideal gas values. This is different from what was reported in Ref.~\cite{qin2012} (cf.\ Sec.~\ref{sec_hs}). However, the experimental points do not follow this trend: they cross the ideal gas values around 0.005 fm$^{-3}$ and are above for lower densities. 
 This also holds for $K_c[d]$, $K_c[h]$, and $K_c[t]$, but for the $\alpha$ particle it is most pronounced. 
The points at low densities occur at very late stages of the HIC where the system has expanded already significantly. Contamination by particle emission from sources other than the coalescing low density gas is the most likely explanation for this. Separation of such contributions is more difficult for late stage lower momenta particles.

In Fig.~\ref{fig_kall_t} we present the ECs as a function of the extracted temperatures. The ideal gas results $K_c^{\rm id}$ are shown by solid lines, to illustrate that these are known functions which solely depend on temperature.
The uncertainty of the temperature determination appears to be more significant than the uncertainty of the density determination; cf.\ Fig.~\ref{fig_kall}. In Fig.~\ref{fig_kall_t}, more of the theoretical data points fall into the constraint region.
In this representation, the lowest temperature data points are also closer to the ideal gas results. 
Despite the different trend of experiment and theory which remains, the ECs from almost all theoretical models are within, or at least close to the experimental constraint at low temperatures.

We emphasize that, since in the ideal case, the ECs are independent of density and only a function of temperature (see Sec.~\ref{sec_definitions}), this representation of the data is perhaps more useful than the one of Fig.~\ref{fig_kall} for low densities and temperatures. For low densities and temperatures, where the ideal case is approached, the extracted density and its uncertainty is almost irrelevant for the ECs. It is gratifying to see, that there is consistency between theory and experiment at lowest temperatures, if the data are presented in this way. At high temperatures, the theoretical models span a broad range of different values for $K_c[\alpha]$, and the experiment shows unambiguously that the simple NSE of ideal gases is ruled out.

Next we have a closer look at $K_c[\alpha]$. LS220, SHT(NL3), and SHO(FSU2.1) show the largest deviations compared to the experiment. For LS220 the apparent underproduction of $\alpha$ particles can be related to the too attractive nucleon interactions at subsaturation densities reported in the literature; see Refs.~\cite{krueger13,fischer14,hempel14}. Furthermore, if we compare with the results of Fig.~\ref{fig_kalpha_3} for different compositions of the HS EOS, the too low values of $K_c[\alpha]$ could possibly be improved by considering light clusters other than $\alpha$ particles. This reduces the partial densities of unbound nucleons, leading to less attractive mean-field effects, which results in an increase of $K_c[\alpha]$.

In the regime of interest, in the SHT(NL3) and SHO(FSU2.1) EOS, 
interactions are only incorporated via the virial coefficients, and
thus there is no suppression of light clusters at high
densities. 
Mean-field interactions of nucleons that are usually fitted to the saturation point of nuclear
matter and to properties of finite nuclei are not incorporated.
This explains the overprediction of $\alpha$ particles. On the other hand
we want to remind the reader that these are the only two EOSs where we
could not implement the constraint $Z\leq2$; see
Sec.~\ref{sec_gshen}. As another aspect, the second virial coefficient
deduced from nucleon-nucleon scattering which is used in these models,
has contributions from two-nucleon bound and continuum states. Here,
these states are not treated as explicit degrees of freedom, but are
attributed to the free nucleon fraction. This treatment represents a
lower limit for $K_c[\alpha]$. The remaining five models, STOS, QS,
gRDF, HS(DD2), and SFHo, are mostly within the error band of $K_c[\alpha]$. STOS is at the lower limit, HS(DD2) and SFHo at the upper. The gRDF and QS models are almost on top of the experimental values. 

Next we discuss the ECs of deuterons, tritons, and helions in more detail. As is clear from the exploration in Sec.~\ref{sec_dofs}, their measurements represent important complementary information, which should not be disregarded. The neglect of these degrees of freedom also affects the EC of $\alpha$ particles. By construction, the first group of models cannot explain this experimental data. It is very satisfactory to see that all of the models from the second group give results within or close to the experimental ranges. 

\section{Summary and Conclusions}
\label{sec_summary}
In this article we investigated cluster formation and medium
modifications in warm nuclear matter at subsaturation densities. 
We constrained a selection of
different SN EOS models by comparing their predictions with
measurements of equilibrium constants (ECs) in HICs, in a similar way as was done in
Ref.~\cite{qin2012} but with an extension to a larger number of light
  cluster species.  Thereby we concentrated on the aspects which were not discussed in full detail in Ref.~\cite{qin2012} and took into account systematic differences between low density matter in SNe and in HICs.
 
First of all we pointed out that the ECs are only independent of
composition, density, and asymmetry for ideal Maxwell-Boltzmann gases without
interactions. In this case they are only a function of
temperature. As soon as interactions and/or Fermi-Dirac or
Bose-Einstein statistics are included, a dependence on
  composition and density arises. Deviations of ECs from the ideal
values measure the strength of interactions. Thus ECs
represent very useful and instructive quantities. Furthermore, some of
the systematic uncertainties are reduced when one uses ECs instead of
particle yields or mass fractions. 

As was illustrated for the HS model, ECs depend on the asymmetry of
the system, or, equivalently, the total proton fraction $Y_p^{\rm tot}$. Therefore we chose
$Y_p^{\rm tot}=0.41$ for all
  theoretical calculations throughout the
paper, corresponding to the value extracted from the experiments for
the emitting source. In SN matter, for which the thermodynamic system
size can be considered as infinite, arbitrarily heavy nuclei can be
formed. In HICs, one has finite size and time limitations. Here we
considered $Z\leq2$ as a constraint for the nuclei which can be formed in the experiment, and applied it in the various theoretical models (where possible). We also removed Coulomb correction for electron screening, which is only relevant for SN matter 
where a background of electrons is present to maintain electric charge neutrality.
For some of the models these modifications lead to better agreement with experiment (e.g., HS(DD2)), for others it got worse (e.g., LS220).

In Ref.~\cite{qin2012}, only the EC of the $\alpha$ particle was
considered for the comparison with EOS predictions. Here we used the
full experimental data set, which also includes measurements of the
ECs of deuterons, tritons, and helions. Some of the considered models
do not include these nuclei as explicit degrees of freedom (LS220,
STOS, SHT(NL3), SHO(FSU2.1)). Such models can obviously not explain
the measured ECs of these particles. 
Even though this is a trivial result, the comparison with the full experimental 
data set gives a more complete picture of the characteristics of the various models. 

We found that all theoretical models approach the ideal gas values at
lowest densities. This is the expected behavior, because there
interactions become weak. Here we corrected some unexpected deviations
which were reported in Ref.~\cite{qin2012}. 
Furthermore, at such conditions a representation of the
ECs as a function of temperature  
is of interest because in the ideal NSE model they are solely functions of $T$. 
When we present the ECs as a function of the extracted temperature rather than density, we also achieve approximate consistency between experiment, the theoretical predictions and the ideal gas values at lowest densities, within or at least close to the experimental error bars. 

In contrast, at highest densities, the experiment shows
unambiguously that an ideal gas description is ruled out, and
that interactions have to be taken into account. Regarding the EC
$K_c[\alpha]$ of the $\alpha$ particle (which is included in all of the
considered models), the LS220, SHO(FSU2.1), and SHT(NL3) show the
largest deviations. In LS220, an underestimation is observed, which we
relate to missing other light nuclei and/or too attractive nucleon
interactions. For the considered temperatures and densities,
SHO(FSU2.1) and SHT(NL3) only include the second-order virial
coefficients for nucleons and $\alpha$ particles in order to account for
  their interaction. This results in an overprediction of
$K_c[\alpha]$. A better agreement could possibly be achieved, if
higher-order virial coefficients or additional light clusters in
  the virial EOS description were included. The former aspect could lead to more binding among the nucleons as observed for the mean-field interactions that are fitted to properties of finite nuclei. Furthermore, the virial EOS has no mechanism of cluster suppression at highest densities, which could also be a contribution to the observed differences. An essential progress is the generalized Beth-Uhlenbeck formula \cite{SRS} which includes also mean-field contributions in a systematic way.

Four of the models which we considered (QS, gRDF, HS(DD2), SFHo) are
fully compatible with the experimental data, or show at least
only minor deviations from the experiment. From the comparison with
the models that fail to explain the full experimental data set, we can
identify the following three ingredients that seem to be
necessary for the description of clusterized nuclear matter at the
densities and temperatures of interest: (i) consideration of all relevant particle degrees of freedom, (ii) mean-field effects of the unbound nucleons, and (iii) a suppression mechanism for bound clusters at high densities. 

Regarding (iii), we compared two different approaches: the classical or phenomenological excluded volume approach (HS(DD2), SFHo) and quasiparticle self-energy shifts based on a quantum mechanical description (QS, gRDF). Only in the latter approaches is there a competition between repulsive Pauli blocking and attractive cluster mean field. 
In principle, the excluded volume concept can be considered as a simple approximation to a full QS treatment similar to the van der Waals EOS; cf.\ Ref.~\cite{hempel11b}.
Unfortunately, the experimental data are not accurate enough to draw firm conclusions about the differences of these approaches, even though they show some sensitivity to the parametrization of the self-energy shifts. 
The major uncertainty in the present investigation does not originate from the experiment itself, but from the density and temperature extraction from the expanding source. For a more refined answer, probably ``densitometers'' and ``thermometers'' have to be used, which include interactions and medium effects in a consistent manner.

Obviously, the quantum mechanical description of medium effects allows
insights which cannot be achieved with phenomenological
approaches. For example the role of continuum correlations 
and how they are modified by the medium can be addressed. 
From a more fundamental point of view, there is no basic definition that distinguishes the contribution of bound states and of the continuum.  Whereas this subdivision is irrelevant for the EOS, it is not clear what happens with the contributions of the continuum in the HIC experiments. For gRDF we have assumed that continuum correlations decay into unbound nucleons,
  simply because they are not bound. In the QS approach, we have separated the mean-field part, and considered only the remaining residual part contributing to the cluster yields. A satisfactory answer can be found only within a nonequilibrium approach to HICs.  
Transport calculations with cluster formation (see, e.g., Refs.~\cite{zubarev96_1,zubarev96_2,kuhrts2001} and references given therein) could help to clarify this interesting issue of how continuum correlations evolve in the expanding system of a HIC, and whether they decay into unbound nucleons or transform into a bound state.

The supernova EOS is a wide field of research where many different aspects of nuclear physics come together. The discussion presented here deals only with one particular but nevertheless important aspect, the formation of light clusters. In supernova matter, heavy nuclei can also be formed and have to be considered in a realistic description, especially for the early phase of the supernova collapse. The experimental results which we use do not allow us to constrain the medium effects on the component of heavy nuclei. These carry uncertainties in the theoretical predictions \cite{buyuk13,aymard2014} similarly large as for light nuclei, which are, however, of a different nature.

Our selection of EOSs for clusterized nuclear matter is by far not complete. Other interesting approaches are for example the statistical multifragmentation model SMM \cite{bondorf95,botvina08} which is well established for the analysis of multifragmentation experiments, but also available as a tabulated SN EOS \cite{buyuk14}. The EOS of Furusawa \textit{et al}.\ \cite{furusawa13} uses Pauli-blocking shifts and excluded volume effects together to dissolve light clusters at high densities. Variations of the gRDF model can be found in Refs.~\cite{ferreira2012,avancini2012}. There are also the statistical models of Ref.~\cite{raduta2010,raduta2014} which have some aspects in common with the HS model. For the virial EOS, there are extensions available which take into account nuclei with $A=2$ and 3 \cite{oconnor07,arcones08}. It would be interesting to repeat our comparison with these models, to validate or refine our main conclusions.

\subsection*{Acknowledgments}
This work was initiated during the ECT* workshop ``Simulating the
Supernova Neutrinosphere with Heavy Ion Collisions'' in April
2014. The authors acknowledge the stimulating and interesting
presentations and discussions during this workshop from all of the
participants and also the support from the ECT*. Partial support comes
from ``NewCompStar,'' COST Action MP1304. M.H.\ is supported by the Swiss
National Science Foundation. S.T.\ acknowledges support by the
Helmholtz Association (HGF) through the Nuclear Astrophysics Virtual 
Institute (NAVI, VH-VI-417). 
J.B.N. acknowledges support from the United States Department of Energy under Grant No.~DE-FG03-93ER40773 and the Robert A. Welch Foundation under Grant No.~A0330.

\appendix

\section{New parametrization of binding energy shifts in the gRDF EOS}
\label{app_grdf}
The calculations of
R\"opke \cite{typel09} suggest a more or less linear dependence on
the effective density for the binding energy shifts of light
nuclei. Thus linear functions are assumed in gRDF as long as the
ground states of the clusters are still bound. For higher densities a
stronger increase of the shifts is used in order to suppress them
there. 
The binding energy shift of a light cluster $i$ with $Z_{i}$ protons and
  $N_{i}$ neutrons is explicitly given by
\begin{equation}
\Delta B_{i}(n_{p}^{\rm tot},n_{n}^{\rm tot},T ) = -
f_{i}(\tilde{n}_{i},n_{i}^{(d)}) \: \delta B_{i}(T)\: n_{i}^{(d)}
\end{equation}
with temperature dependent quantities
$\delta B_{i}(T)$, which are identical to those in
Ref.~\cite{typel09}, and shift functions $f_{i}$. They depend on the
effective density
\begin{equation}
 \tilde{n}_{i} = \frac{2}{Z_{i}+N_{i}} \left[ Z_{i} n_{p}^{\rm tot} +
   N_{i} n_{n}^{\rm tot} \right]
\end{equation}
and the dissociation density 
\begin{equation}
 n_{i}^{(d)} =    \frac{B_{i}^{0}}{\delta B_{i}(T)}
\end{equation}
with saturation density $n_{\rm sat}$ of the DD2 parametrization 
and the experimental vacuum binding
  energy $B_{i}^{0}$. The quantities $n_{p}^{\rm tot}$ and $n_{n}^{\rm
  tot}$ are the total proton and neutron densities, respectively,
including nucleons that are free or bound in nuclei.
Introducing $x=\tilde{n}_{i}/n_{i}^{(d)}$ and
 $y=n_{\rm sat}/n_{i}^{(d)}$, the shift function reads
\begin{equation}
 f_{i} = \left\{ 
 \begin{array}{lll}
 x & \mbox{if} & x \leq 1 \\
 x + \frac{(x-1)^{3}(y-1)}{3(y-x)} & \mbox{if} & x > 1 \land x < y
 \end{array} \right. \: .
\end{equation}
In the previous parametrization given in Ref.~\cite{typel09}, 
$f_{i}=x+x^{2}/2$ is used throughout.
For $x \to y$ the binding energy shift diverges and for
$\tilde{n}_{i} \geq n_{i}^{(d)}$ the cluster does not appear any more.

\section{QS approach to the contribution of the continuum to the EOS}
\label{app:QS}
A systematic approach to the EOS can be given by a Green-function approach \cite{RMSa,RMSb,cmf2}, and different approximations can be performed for the spectral function and the self-energy, in particular the cluster decomposition and the quasiparticle concept; see Ref.~\cite{roepke2014b} and further references given there.
We give here some forthcoming results used in the present work.

For the $A$-nucleon cluster, the  in-medium Schr\"odinger equation 
\begin{widetext} 
\begin{eqnarray}
&&[E_1(1;T,\mu_n,\mu_p)+\dots + E_1(A;T,\mu_n,\mu_p) - E_{A \nu}(P;T,\mu_n,\mu_p)]\psi_{A \nu P}(1\dots A)
\nonumber \\ &&
+\sum_{1'\dots A'}\sum_{i<j}[1-n(i;T,\mu_n,\mu_p)- n(j;T,\mu_n,\mu_p)]V(ij,i'j')\prod_{k \neq 
  i,j} \delta_{kk'}\psi_{A \nu P}(1'\dots A')=0\,
\label{waveA}
\end{eqnarray}
\end{widetext}
is derived from the Green-function approach.
This equation contains the effects of the medium in the single-nucleon quasiparticle shift 
$\Delta E_{\tau_1}^{\rm SE}(p_1;T,\mu_n,\mu_p) =  E_1(1;T,\mu_n,\mu_p)-(\hbar^2p_1^2/2 m)$
as well as in the Pauli blocking terms given by the occupation numbers $n(1;T,\mu_n,\mu_p)$
in the phase space of single-nucleon states $|1 \rangle$.  Thus, two effects have to be considered: the quasiparticle
energy shift and the Pauli blocking. 
The single nucleon quasiparticle shifts is approximated, for instance, by appropriate parametrizations of the RMF such as DD2 \cite{typel09}.
The self-energy and Pauli blocking shifts for the light elements are obtained from the in-medium Schr\"odingier equation (\ref{waveA}).
Explicit expressions that approximatively describe the shift of the light element quasiparticles ($A \leq 4$) are given in Ref.~\cite{roepke11a}.

The summation over the internal quantum number includes the continuum contribution $v_c$. They give a contribution to the EOS, as clearly shown
by the Beth-Uhlenbeck formula for the second virial coefficient. We will not discuss here the different ways to introduce it into the EOS, 
see \cite{SRS,voskresenskaya2012,roepke2014a,roepke2014b}, but only underline that it is relevant for the contributions of the two-nucleon (deuteron) system because of the small binding energy $B_d=-E^{(0)}_d = 2.225$ MeV which is of the order (or smaller) compared with $T$. For the stronger bound clusters ($t,h,\alpha$), 
the continuum states are of less relevance.

In addition to Ref.~\cite{roepke2014b} where only the medium modified continuum states with $P=0$ have been considered, we consider the residual continuum contributions as a function of $P$. Like the in-medium shift of the binding energy, the medium modification of scattering states is strongly depending on $P$ because the Pauli blocking changes quickly with the c.o.m.\ momentum $P$. At high values of $P$ the Pauli blocking and correspondingly the reduction of the contributions of the clusters becomes inactive. 

Using the same method given in \cite{roepke2014b} for $P=0$, i.e., scaling the reduction of the binding energy with increasing $n_B$ 
with the known value for $n^{\rm Mott}_c(P,T,Y_p)$, we find for arbitrary $P$ the partial densities $n_c(T,n_B,Y_p)$.
Exemplarily we give the expression for the $\alpha$ particle 
\begin{widetext} 
\begin{eqnarray}
&&n_\alpha(T,n_B,Y_p)=g_\alpha\int dP\frac{1}{2 \pi^2}P^2 
e^{[-2  E_n(P/4,T,n_B,Y_p)-2  E_p(P/4,T,n_B,Y_p)+2 \mu_n+2 \mu_p]/T}\nonumber \\
&&\times  \left\{ (e^{[B_\alpha-\Delta E_\alpha(P,T,n_B,Y_p)]/T}-1) \Theta(B_\alpha-\Delta E_\alpha(P,T,n_B,Y_p))  +v_\alpha(P,T,n_B,Y_p)\right.\nonumber \\
&&\left.+(e^{[B_{\alpha'}-\Delta E_\alpha(P,T,n_B,Y_p)]/T}-1) \Theta(B_{\alpha'}-\Delta E_\alpha(P,T,n_B,Y_p)) +v_\alpha(P,T,n_B,Y_p) \right\} ,
\end{eqnarray}
where $B_\alpha=-E_\alpha^{(0)}\simeq 28.3$~MeV is the binding energy of the ground state and $B_{\alpha'}\simeq 8.1$~MeV of the excited state. $g_\alpha=1$ is the spin degeneracy factor. Analog expressions hold for $n_d$, $n_t$, and $n_h$ but with only one bound state $E_i^{(0)}$.
The residual contribution of the continuum correlations is approximated as
\begin{equation}
v_i(P,T,n_B,Y_p)\approx \left\{1.24+\left[ \frac{1}{v^{(0)}(T)}-1.24\right]e^{\gamma_i(P,T,n_B,Y_p) n_B/T_{\rm eff}(T,n_B)}\right\}^{-1},
\end{equation}
with
\begin{eqnarray}
\gamma_d(P,T,n_B,Y_p) &=&1873.24\, \exp\left[-P^2{\rm fm}^{2}/(1.84632 + 0.161695 T{\rm MeV}^{-1} + \,\,\,0.17266 P^2{\rm fm}^{2})\right]{\rm MeV \ fm}^{3}, \nonumber \\
\gamma_t(P,T,n_B,Y_p) &=&2773.22 \, \exp\left[-P^2{\rm fm}^{2}/(4.66711 + \quad0.3037 T{\rm MeV}^{-1} + \quad0.1439 P^2{\rm fm}^{2})\right]{\rm MeV \ fm}^{3}, \nonumber \\
\gamma_h(P,T,n_B,Y_p) &=&2843.52 \, \exp\left[-P^2{\rm fm}^{2}/(4.67929 + 0.284855 T{\rm MeV}^{-1} + \,\,\,0.14037 P^2{\rm fm}^{2})\right]{\rm MeV \ fm}^{3} ,\nonumber \\
\gamma_\alpha(P,T,n_B,Y_p) &=&3268.84\, \exp\left[-P^2{\rm fm}^{2}/(9.75141 + 0.692198 T{\rm MeV}^{-1} + 0.243567 P^2{\rm fm}^{2})\right]{\rm MeV \ fm}^{3}\,,
\end{eqnarray}
\end{widetext} 
and
\begin{equation}
T_{\rm eff}(T,n_B)=5.5 {\rm MeV}+0.5 \,T +60\, n_B {\rm MeV\, fm}^3\,.
\end{equation}
The residual virial coefficient at zero density \cite{roepke2014b} coincides well with the empirical data for the full virial coefficient given in Ref.~\cite{horowitz06b},
\begin{equation}
v^{(0)}(T) = 0.30857 + 0.65327 \,e^{-0.102424 T/{\rm MeV}} \, .
\end{equation}

The residual contributions of the continuum avoid double counting because the mean-field energies arise from the interaction of continuum states.
The problem is that the RMF approach is introduced semiempirically by fitting data in the region of saturation density. More consistent would be a readjustment of the parameter values, e.g., with microscopic Dirac-Brueckner calculations.

The self-consistent treatment of correlations in the medium demands further work, beyond the introduction of an effective chemical potential and effective temperature to calculate the Pauli blocking. The systematic inclusion of correlations in nuclear matter should include also the self-consistent treatment of cluster formation in the self-energy
as well as in the Pauli blocking term. For discussion see Ref.~\cite{roepke2014b} where an approximate approach is given. 

\bibliographystyle{apsrev}
\bibliography{literature}

\end{document}